\documentclass[aps,pra,twocolumn,groupedaddress,showpacs,superscriptaddress,scrartcl]{revtex4-1}
\usepackage{amssymb,amsmath,graphicx,epstopdf,hyperref,amsthm,dsfont,enumerate,bm,color}
\newtheorem{lemma}{Lemma}

\begin{document}
\interfootnotelinepenalty=10000

\title{One-out-of-$m$ spacetime{-}constrained oblivious transfer}

\author{Dami\'an Pital\'ua-Garc\'ia}
\email{D.Pitalua-Garcia@damtp.cam.ac.uk}
\affiliation{Centre for Quantum Information and Foundations, DAMTP, Centre for Mathematical Sciences, University of Cambridge, Wilberforce Road, Cambridge, CB3 0WA, U.K.}

\date{\today}

\begin{abstract}
In one-out-of-$m$ spacetime-constrained oblivious transfer (SCOT), Alice and Bob agree on $m$ pairwise spacelike separated output spacetime regions $R_0,R_1,\ldots, R_{m-1}$ in an agreed reference frame in a spacetime that is Minkowski, or close to Minkowski; Alice inputs a message $\bold{x}_i$ in the causal past of a spacetime point $Q_i$ of $R_i$, for $i\in\{0,1,\ldots,m-1\}$; Bob inputs $b\in\{0,1,\ldots,m-1\}$ in the intersection of the causal pasts of $Q_0,Q_1,\ldots,Q_{m-1}$ and outputs $\bold{x}_b$ in $R_b$; Alice remains oblivious to $b$ anywhere in spacetime; and Bob is unable to obtain $\bold{x}_i$ in $R_i$ and $\bold{x}_j$ in $R_j$ for any pair of different numbers $i,j\in\{0,1,\ldots,m-1\}$. We introduce unconditionally secure one-out-of-$m$ SCOT protocols extending the one-out-of-two SCOT protocols of Refs. \cite{PG15.1,PGK18.2}, for arbitrary integers $m\geq 2$. We define the task of one-out-of-$m$ distributed quantum access with classical memory (DQACM), which works as a subroutine to implement a class $\mathcal{P}_{\text{CC}}$ of one-out-of-$m$ SCOT protocols where distant agents only need to communicate classically. We present unconditionally secure one-out-of-$m$ DQACM protocols and one-out-of-$m$ SCOT protocols of the class $\mathcal{P}_{\text{CC}}$, for arbitrary integers $m\geq2$. 
We discuss various generalizations of SCOT. In particular, we introduce a straightforward extension of SCOT to a $k$-out-of-$m$ setting, and suggest protocols where distant agents only need to communicate classically, while we leave the investigation of their security as an open problem.
\end{abstract}


\maketitle

\section{Introduction}

One-out-of-$m$ oblivious transfer is a fundamental cryptographic task
that works as a primitive in secure computation \cite{K88}.  Secure computation \cite{Y82} is an area of cryptography in which  two or more mistrustful parties compute a joint function of their private inputs in such a way that there is no information about their inputs, which does not follow from the output of the computation, revealed to the other parties. In an one-out-of-$m$ oblivious transfer protocol, Alice inputs $m\geq 2$ messages $\bold{x}_0,\bold{x}_1,\ldots,\bold{x}_{m-1}$, Bob inputs a number $b\in\{0,1,\ldots,m-1\}$, and Bob outputs the message $\bold{x}_b$. An one-out-of-$m$ oblivious transfer protocol must satisfy two security conditions: 1) the condition of security for honest Alice, also denoted as security against dishonest Bob, according to which if Alice follows the protocol honestly but Bob does not, Bob cannot obtain more than one of Alice's messages; and 2) the condition of security for honest Bob, also denoted as security against dishonest Alice, according to which if Bob follows the protocol honestly but Alice does not, Alice remains oblivious to Bob's input $b$.

Protocols for one-out-of-$m$ oblivious transfer and more general secure computations have been proposed, with the security being based in computational or technological assumptions, for example, the assumed difficulty of finding the prime factors of large integers \cite{Y86,K88}, or the assumption that quantum memories are bounded or noisy \cite{WST08}. However, one-out-of-$m$ oblivious transfer and more general secure computations cannot be implemented with unconditional security in the standard setting of non-relativistic quantum cryptography \cite{L97,C07,BCS12}, i.e. it is impossible to guarantee their security only from the laws of quantum physics. In particular, Lo's no-go theorem \cite{L97} states that if a protocol for one-out-of-$m$ oblivious transfer is unconditionally secure against dishonest Alice then, with sufficiently advanced quantum technology, Bob can obtain all messages $\bold{x}_0,\bold{x}_1,\ldots,\bold{x}_{m-1}$.

One-out-of-$m$ oblivious transfer and more general secure computations cannot achieve unconditional security even in the more general setting of relativistic quantum cryptography, introduced by Kent \cite{K99,K12.1}, in which each party in the protocol has trusted agents performing quantum computations and communications at different spacetime points, some of which are spacelike separated. For example, in the case of one-out-of-$m$ oblivious transfer, if a quantum relativistic protocol taking place in a finite region $R$ of spacetime, in some reference frame $\mathcal{F}$, is unconditionally secure against Alice, then it follows from Lo's no-go theorem that Bob can perform the protocol honestly with an input $b=0$ and obtain $\bold{x}_0$ in $R$ and then apply a unitary operation $U$ on his global quantum system, which is spread among various locations, and complete it within the spacetime region $T$ consisting in the intersection of the causal futures of all the spacetime points of $R$ -- for example, Bob can simply send all his quantum systems to a common spacetime point within $T$ and then apply $U$ there -- and then apply a quantum measurement to obtain $\bold{x}_1$ and then proceed similarly to obtain $\bold{x}_2$ and so on. This is in contrast to other tasks in mistrustful cryptography, like coin tossing and bit commitment, for which unconditionally secure protocols cannot exist in non-relativistic quantum cryptography \cite{LC98,LC97,M97}, but for which there are unconditionally secure protocols in relativistic quantum cryptography \cite{K99.2,K99,K05.2,K11.2,K12,LKBHTWZ15}.

Nevertheless, two relativistic variations of one-out-of-two oblivious transfer have been recently proposed, denoted as location-oblivious data transfer (LODT) \cite{K11.3} and spacetime-constrained oblivious transfer (SCOT) \cite{PG15.1}, which have been shown to achieve unconditional security \cite{K11.3,PG15.1}. In LODT, Alice transfers a message to Bob at a random location in spacetime that neither Alice nor Bob can determine in advance, and Alice remains oblivious to the location where Bob received the message. In SCOT, according to a bit $b$ input by Bob, Bob either obtains a message $\bold{x}_0$ of Alice in a spacetime region $R_0$ or a message $\bold{x}_1$ of Alice in a spacetime region $R_1$, where $R_0$ and $R_1$ are spacelike separated, and where Alice remains oblivious to Bob's input $b$. Interestingly, LODT and SCOT are the only known cryptographic tasks that necessitate both the no-superluminal principle of relativity theory and the properties of quantum information to achieve unconditional security \cite{K11.3,PG15.1}, in contrast to coin tossing and bit commitment, for example, for which there are unconditionally secure relativistic protocols that are purely classical \cite{K99.2,K99,K05.2,LKBHTWZ15}.

Two unconditionally secure protocols for SCOT have been presented in the academic literature \cite{PG15.1, PGK18.2}. The protocol of Ref. \cite{PG15.1} requires the preparation of random Bennett-Brassard 1984 (BB84) \cite{BB84} states and their secure transmission to distant laboratories. The protocol of Ref. \cite{PGK18.2} requires the preparation and transmission of quantum states between adjacent laboratories, and the transmission of classical information to distant laboratories. Here, we introduce unconditionally secure protocols that generalize those of Refs. \cite{PG15.1, PGK18.2} to the one-out-of-$m$ setting, in which Alice inputs $m\geq2$ messages $\bold{x}_0,\bold{x}_1,\ldots,\bold{x}_{m-1}$, there are $m\geq 2$ pairwise spacelike separated output spacetime regions $R_0,R_1,\ldots,R_{m-1}$, Bob inputs an integer $b\in\{0,1,\ldots,m-1\}$ and outputs $\bold{x}_b$ in $R_b$, Alice remains oblivious to $b$ anywhere in spacetime, and Bob cannot output $\bold{x}_i$ in $R_i$ and also $\bold{x}_j$ in $R_j$ for any pair of different numbers $i,j$ from the set $\{0,1,\ldots,m-1\}$.

Potential applications of one-out-of-$m$ SCOT include situations where Bob needs to access, 
at a specific location and within a short interval of time, one of various pieces of information input by Alice, and Bob requires his choice of accessed piece of information to remain secret to Alice. For example, potential applications of one-out-of-$m$ SCOT and generalizations are in high frequency trading strategies (HFT) in the stock market, where must transaction are completed within half a millisecond \cite{WF10}. Consider for example in this case the following situation. Alice is a company that sells information about the stock market in real time in a set of different possible locations and Bob is a company that trades in the stock market using HFT strategies. Alice offers Bob one piece of her database $\bold{x}_0,\bold{x}_1,\ldots,\bold{x}_{m-1}$, each being information on the stock market at the respective location $\text{L}_0,\text{L}_1,\ldots,\text{L}_{m-1}$ at real time. Each $\text{L}_i$ could be the location of a stock market in some part of the world, for instance New York, Toronto, Paris, London, Tokyo, etc. Bob pays Alice a fixed amount of money to obtain an entry $\bold{x}_b$ in the location $\text{L}_b$ in real time. Bob requires that his choice $b$ remains private from Alice, while Alice requires that Bob cannot access her entry $\bold{x}_i$ in $\text{L}_i$ at real time, for more than one $i$ from the set $\{0,1,\ldots,m-1\}$. One-out-of-$m$ SCOT guarantees with unconditional security that Alice cannot learn Bob's choice $b$ anywhere in spacetime and that within a time interval smaller than 0.5 ms, which is relevant for HFT strategies, Bob cannot obtain $\bold{x}_i$ in $\text{L}_i$ and also $\bold{x}_j$ in $\text{L}_j$ for any pair of different numbers $i,j$ from the set $\{0,1,\ldots,m-1\}$ if the distance between any pair of locations from the set $\text{L}_0,\text{L}_1,\ldots,\text{L}_{m-1}$ is at least 150 km, which is the maximum distance that light can travel in 0.5 ms in the approximately Minkowski spacetime near the Earth surface.

We mainly focus on a class $\mathcal{P}_{\text{CC}}$ of one-out-of-$m$ SCOT protocols extending those of Ref. \cite{PGK18.2}, which require classical -- but not quantum -- communication among distant locations, and we show them unconditionally secure. We introduce a quantum-cryptography task denoted as one-out-of-$m$ distributed quantum access with classical memory (DQACM), which works as a fundamental primitive to construct the class $\mathcal{P}_{\text{CC}}$ of one-out-of-$m$ SCOT protocols.

Broadly speaking, a protocol for one-out-of-$m$ DQACM consists in the following steps. Alice encodes $m$ messages $\bold{r}_0,\bold{r}_1,\ldots,\bold{r}_{m-1}$ chosen randomly from predetermined sets in a quantum state $\lvert \Psi_{\bold{r}}^{\bold{s}}\rangle$, where $\bold{r}=(\bold{r}_0,\bold{r}_1,\ldots,\bold{r}_{m-1})$, and where $\bold{s}$ denotes a basis chosen randomly by Alice from a predetermined set of non-mutually orthogonal bases. Alice sends $\lvert \Psi_{\bold{r}}^{\bold{s}}\rangle$ to Bob. Bob chooses a random number $c\in\{0,1,\ldots,m-1\}$, applies a quantum measurement $\text{M}_c$ on $\lvert \Psi_{\bold{r}}^{\bold{s}}\rangle$, and obtains a classical measurement outcome $\bold{d}$. At a later time Alice gives $\bold{s}$ to Bob, who then uses $\bold{d}$ and $\bold{s}$ to learn Alice's input $\bold{r}_c$. A one-out-of-$m$ DQACM protocol must satisfy a security condition against dishonest Bob, according to which, if Alice follows the protocol honestly and Bob follows a cheating strategy in which he applies an arbitrary quantum operation $O$ on $\lvert \Psi_{\bold{r}}^{\bold{s}}\rangle$ that produces at least two quantum systems $B_0$ and $B_1$, and then quantum measurements $\tilde{\text{M}}_0^{\bold{s}}$ and $\tilde{\text{M}}_1^{\bold{s}}$ are applied on $B_0$ and $B_1$ after receiving $\bold{s}$, giving classical outcomes $\bold{r}_i'$ and $\bold{r}_j'$, respectively, then the probability that $\bold{r}_i'$ equals $\bold{r}_i$ and $\bold{r}_j'$ equals $\bold{r}_j$ is negligible for Alice's input messages of large size, and for any pair of different numbers $i,j$ from the set $\{0,1,\ldots,m-1\}$.

We introduce a class $\mathcal{C}$ of one-out-of-$m$ DQACM protocols. We show that by satisfying a few properties the protocols of this class are unconditionally secure. We give specific examples of unconditionally secure one-out-of-$m$ DQACM protocols from the class $\mathcal{C}$.

We also briefly discuss various generalizations of one-out-of-$m$ SCOT. In particular, we suggest a definition for $k$-out-of-$m$ SCOT, for arbitrary natural numbers $k<m$ and $m\geq 2$, according to which Alice and Bob agree on $m$ pairwise spacelike separated output spacetime regions $R_0,R_1,\ldots,R_{m-1}$, Alice inputs $m$ messages $\bold{x}_0,\bold{x}_1,\ldots,\bold{x}_{m-1}$, Bob obtains Alice's input $\bold{x}_{b_l}$ in the output spacetime region $R_{b_l}$, for $k$ different numbers $b_l\in\{0,1,\ldots,m-1\}$ chosen by Bob, Bob cannot obtain $\bold{x}_i$ in $R_i$, for more than $k$ different elements $i$ from the set $\{0,1,\ldots,m-1\}$, and Alice remains oblivious to Bob's inputs $b_l$, anywhere in spacetime. We suggest protocols for $k$-out-of-$m$ SCOT where communication between distant locations is only classical, based on the primitive of $k$-out-of-$m$ DQACM, which is a natural generalization of one-out-of-$m$ DQACM into a $k$-out-of-$m$ setting, but we leave as an open question to investigate whether they are unconditionally secure. In particular, we propose protocols for $k$-out-of-$m$ DQACM extending the class $\mathcal{C}$ of protocols for one-out-of-$m$ DQACM, and we leave as an open problem to show whether they are unconditionally secure.


This paper is organized as follows. In section \ref{section2}, we describe the setting of relativistic quantum cryptography, we provide some mathematical notation and we recall the SCOT protocols of Refs. \cite{PG15.1,PGK18.2}. We define one-out-of-$m$ SCOT in section \ref{section3}. In section \ref{section4}, we introduce an unconditionally secure one-out-of-$m$ SCOT protocol $\mathcal{P}_{\text{QC}}$ that extends the one-out-of-two SCOT protocol of Ref. \cite{PG15.1}, and which requires quantum communication between distant locations. We define the task of one-out-of-$m$ DQACM in section \ref{section5}. In section \ref{section6}, we present a class $\mathcal{C}$ of unconditionally secure protocols for one-out-of-$m$ DQACM,  we show that this class of protocols are unconditionally secure from the satisfaction of a few properties, and we give specific examples of this class of protocols. In section \ref{section7}, we introduce a class $\mathcal{P}_{\text{CC}}$ of one-out-of-$m$ SCOT protocols where communication between distant locations is only classical, and where one-out-of-$m$ DQACM acts as a subroutine; we show the class $\mathcal{P}_{\text{CC}}$ is unconditionally secure if the one-out-of-$m$ DQACM subroutine is unconditionally secure; and we discuss specific examples of protocols from the class $\mathcal{P}_{\text{CC}}$ where the one-out-of-$m$ DQACM subroutine belongs to the class $\mathcal{C}$. Section \ref{section8} discusses generalizations of one-out-of-$m$ SCOT; in particular, definitions of $k$-out-of-$m$ SCOT and $k$-out-of-$m$ DQACM are suggested; protocols for $k$-out-of-$m$ DQACM and $k$-out-of-$m$ SCOT are outlined, with the investigation of their security being left as an open problem.
We conclude in section \ref{section9} with a discussion of our results and of possible connections with other research problems in quantum information and relativistic quantum cryptography.


\section{Preliminaries}
\label{section2}

\subsection{Relativistic quantum cryptography}

In relativistic quantum cryptography, security is guaranteed from 1) the no-superluminal principle of relativity theory, stating that physical systems and information cannot travel faster than light, which is satisfied by quantum theory; and 2) the properties of quantum information, for example, the no-cloning theorem \cite{D82,WZ82}, the impossibility of perfectly distingushing non-orthogonal quantum states, the monogamy of quantum entanglement \cite {T04}, the existence of quantum correlations that violate Bell inequalities \cite{Bell}, etc.

Relativistic-quantum cryptography is usually considered for spacetimes that are Minkowski, or close to Minkowski,  as near the Earth surface. But, relativistic quantum cryptography can also apply to arbitrary curved spacetimes with well defined causal structure, if the parties participating in the cryptographic tasks have a well description of the spacetime geometry, if they cannot substantially alter the geometry of spacetime, and if within the region of spacetime where the cryptographic tasks take place, there are not wormholes or other mechanisms allowing them to send signals faster than the speed of light \cite{K99}.

In the setting of relativistic quantum cryptography, the parties participating in the cryptographic tasks, e.g. Alice and Bob, consist of various agents who process and communicate classical and quantum information at various locations in spacetime.
In general, in a protocol for relativistic quantum cryptography, the participating parties
must agree on spacetime regions where they should communicate classical or quantum information to each other. For this reason, the parties agree on a reference frame $\mathcal{F}$ with global spacetime coordinates $(t,x,y,z)$, where the first entry is temporal and the others are spatial, and where without loss of generality we use units in which the speed of light is unity. In the case of mistrustful cryptography, which includes the task of SCOT considered in this paper, Alice's (Bob's) agents work in collaboration and trust each other, but Alice's agents are mistrustful of Bob's agents and vice versa.

\subsection{Notation}

We define the sets $\text{I}_m=\{0,1,\ldots,m-1\}$ and $[n]=\{1,2,\ldots,n\}$ for any integer numbers $m\geq 2$ and $n\geq 1$. For a string $\bold{a}$ of $n$ entries, we denote the $j$th entry by $a^j$, for $j\in [n]$. The Hamming distance between strings of bits $\bold{a}$ and $\bold{b}$ is denoted by $d(\bold{a},\bold{b})$. The Hamming weight of a string of bits $\bold{a}$ is denoted by $w(\bold{a})$. When applied to bits (bit strings) $\oplus$ denotes (bitwise) sum modulo 2. We denote the complement of a bit $a$ by $\bar{a}=a\oplus 1$, and of a bit $a^j$ by $\bar{a}^j$. The binary entropy of $\gamma\in(0,1)$ is given by $h(\gamma)\equiv -\gamma\log_2 \gamma-(1-\gamma)\log_2 (1-\gamma)$, and of $\gamma\in\{0,1\}$ is defined as zero. We use the following notation for the BB84 states: $\lvert \psi_0^0\rangle=\lvert 0\rangle$, $\lvert \psi_1^0\rangle=\lvert 1\rangle$, $\lvert \psi_0^1\rangle=\lvert +\rangle$, $\lvert \psi_1^1\rangle=\lvert -\rangle$, where $\lvert \pm\rangle = \frac{1}{\sqrt{2}}\bigl(\lvert 0\rangle\pm \lvert 1\rangle\bigr)$. The computational and Hadamard bases are denoted by $\{\lvert 0\rangle,\lvert 1\rangle\}$ and $\{\lvert +\rangle,\lvert -\rangle\}$, respectively.

\subsection{The SCOT protocols of Refs. \cite{PG15.1,PGK18.2}}

It is useful to recall the SCOT protocols of Refs. \cite{PG15.1,PGK18.2}, because we will extend these in following sections. We describe the common setting of these protocols before presenting them.

In the SCOT protocols of Refs. \cite{PG15.1,PGK18.2}, Alice has three agents $\mathcal{A}$, $\mathcal{A}_0$, and $\mathcal{A}_1$; and Bob has three agents $\mathcal{B}$, $\mathcal{B}_0$, and $\mathcal{B}_1$. Alice and Bob agree in a reference frame $\mathcal{F}$ and in two spacelike separated output spacetime regions $R_0$ and $R_1$. Each of Alice's (Bob's) agents controls a secure laboratory. It is useful to consider that the agents $\mathcal{A}$ and $\mathcal{B}$ have adjacent laboratories, and that the agents $\mathcal{A}_i$ and $\mathcal{B}_i$ have adjacent laboratories, for $i\in\{0,1\}$. There is a quantum channel between $\mathcal{A}$ and $\mathcal{B}$. There is a classical channel between $\mathcal{A}_i$ and $\mathcal{B}_i$, for $i\in\{0,1\}$.

In the SCOT protocol of Ref. \cite{PG15.1}, Alice's agents share secure and authenticated classical channels, and Bob's agents share secure and authenticated quantum channels. On the other hand, in the SCOT protocol of Ref. \cite{PGK18.2}, Alice's agents share secure and authenticated classical channels, and Bob's agents share secure and authenticated classical channels, but Bob's agents do not need to share quantum channels; additionally, there is a classical channel, as well as a quantum channel, between $\mathcal{A}$ and $\mathcal{B}$.

\begin{figure}
\includegraphics{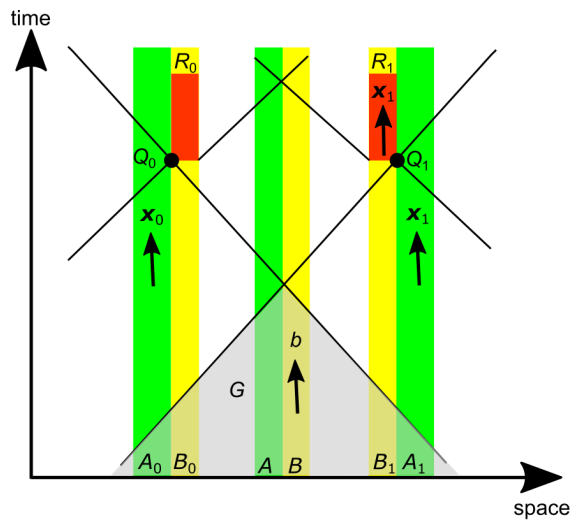}
 \caption{\label{fig1} Illustration of a setting for the SCOT protocols of Refs. \cite{PG15.1, PGK18.2} in a two-dimensional spacetime diagram, in a frame $\mathcal{F}$, of Minkowski spacetime. The world lines of the laboratories of Alice's agents $\mathcal{A}$, $\mathcal{A}_0$, $\mathcal{A}_1$ (green rectangles), and of the laboratories of Bob's agents $\mathcal{B}$, $\mathcal{B}_0$, $\mathcal{B}_1$ (yellow rectangles) are indicated. The small dots represent the spacetime points $Q_0$ and $Q_1$. The thin diagonal lines represent light rays. The spacetime region $G$, consisting in the intersection of the causal pasts of $Q_0$ and $Q_1$, is represented by the grey shaded area. The spacetime regions $R_i$, where Bob's agents must obtain Alice's inputs $\bold{x}_i$, correspond to the small red rectangles, for $i\in\{0,1\}$. We note that $R_0$ and $R_1$ are spacelike separated. Alice inputs a message $\bold{x}_i$ in the causal past of $Q_i$, for $i\in\{0,1\}$. Bob inputs a bit $b$ in $G$ and outputs $\bold{x}_b$ in $R_b$. The case $b=1$ is illustrated.}
\end{figure}

In both protocols, Alice and Bob agree on spacetime points $Q_0$ in $R_0$ and $Q_1$ in $R_1$. We define $G$ as the intersection of the causal pasts of $Q_0$ and $Q_1$. In the notation of Refs. \cite{PG15.1, PGK18.2}, $G$ is the causal past of a spacetime point $P$, which is in the causal past of a spacetime point of $R_0$ and a spacetime point of $R_1$. Alice inputs a $n-$bit string $\bold{x}_i$ in the causal past of $Q_i$, for $i\in\{0,1\}$. Bob inputs a bit $b$ in $G$ and outputs $\bold{x}_b$ in $R_b$. An example of a setting for the SCOT protocols of Refs. \cite{PG15.1, PGK18.2} is given in Fig. \ref{fig1}.

As shown in Refs. \cite{PG15.1,PGK18.2}, both protocols are unconditionally secure. The protocols are trivially unconditionally secure against dishonest Alice, as Alice does not receive any information from Bob. The protocols are unconditionally secure against dishonest Bob: in any cheating strategy by Bob allowed by quantum theory and relativity, the probability $p_n$ that Bob outputs $\bold{x}_0$ in $R_0$ and $\bold{x}_1$ in $R_1$ decreases exponentially with $n$, satisfying $p_n\leq \bigl(\frac{1}{2}+\frac{1}{2\sqrt{2}}\bigr)^n$. The protocols can be extended to allow a small fraction of errors in Bob's output message, while still satisfying unconditional security for dishonest Bob.

\subsubsection{The SCOT protocol of Ref. \cite{PG15.1}}

\begin{enumerate}
\item Agent $\mathcal{A}$ encodes a random $n-$bit string $\bold{r}$ in a quantum state $\lvert \psi_{\bold{r}}^{\bold{s}}\rangle=\bigotimes_{j\in[n]}\lvert \psi_{r^j}^{s^j}\rangle$ of $n$ BB84 states, where the $n-$bit string $\bold{s}$ is random and denotes the bases. $\mathcal{A}$ sends $\lvert \psi_{\bold{r}}^{\bold{s}}\rangle$ to $\mathcal{B}$, who receives it in $G$. 

\item $\mathcal{B}$ obtains his input bit $b$ in $G$ and redirects the received state $\lvert \psi_{\bold{r}}^{\bold{s}}\rangle$ to his colleague $\mathcal{B}_b$, who receives it in the causal past of at least one spacetime point of $R_b$.

\item For $i\in\{0,1\}$, $\mathcal{A}$ sends copies of $\bold{s}$ and $\bold{r}$ to her colleague $\mathcal{A}_i$, who receives them in the causal past of $Q_i$. 

\item For $i\in\{0,1\}$, $\mathcal{A}_i$ obtains her input message $\bold{x}_i\in\{0,1\}^n$ in the causal past of $Q_i$.

\item For $i\in\{0,1\}$, $\mathcal{A}_i$ gives $\bold{s}$ and $\bold{t}_i=\bold{x}_i\oplus \bold{r}$ to $\mathcal{B}_i$ at the spacetime point $Q_i$.

\item $\mathcal{B}_b$ measures the quantum state $\lvert \psi_{\bold{r}}^{\bold{s}}\rangle$ in the basis labeled by $\bold{s}$ and obtains the encoded string $\bold{r}$ in $R_b$.

\item $\mathcal{B}_b$ computes the message $\bold{x}_b=\bold{r}\oplus \bold{t}_b$ and outputs it in $R_b$.

\end{enumerate}

\subsubsection{The SCOT protocol of Ref. \cite{PGK18.2}}

This SCOT protocol consists of two main stages. Stage I includes quantum communication between agents $\mathcal{A}$ and $\mathcal{B}$, which can take place within their adjacent laboratories, and which can take an arbitrarilly long time, but which must be completed within $G$. For $i\in\{0,1\}$, stage II includes fast classical processing and communication between the agents $\mathcal{A}_i$ and $\mathcal{B}_i$, which again can take place within their adjacent laboratories; it also includes classical communication between the -- possibly distant -- pairs of agents $\mathcal{A}$ and $\mathcal{A}_i$, and $\mathcal{B}$ and $\mathcal{B}_i$. Steps 1 to 6 take place within $G$. 

\begin{center}
\emph{Stage I}
\end{center}

\begin{enumerate}
\item For $i \in \{0,1\}$, $\mathcal{A}$ generates  random $n-$bit strings  $\bold{r}_0,\bold{r}_1,\bold{s}$ and sends copies to $\mathcal{A}_i$, who receives them in the causal past of $Q_i$.

\item $\mathcal{A}$ prepares a system $A=A^1A^2\cdots A^n$ of $n$ qubit-pairs $A^1=A_0^1A_1^1, \ldots, A^n=A_0^nA_1^n$ in the quantum state $\lvert \psi_{\bold{r}_0,\bold{r}_1}^{\bold{s}}\rangle=\bigotimes_{j\in[n]}\bigl\lvert \psi_{r_0^jr_1^j}^{s^j}\bigr\rangle_{A_0^jA_1^j}$, where
\begin{equation}
\label{1}
\bigl\lvert \psi_{r_0^jr_1^j}^{s^j}\bigr\rangle_{A_0^jA_1^j}=\bigl\lvert \psi_{r_0^j}^0\bigr\rangle_{A_{s^j}^j} \bigotimes  \bigl \lvert \psi_{r_1^j}^1\bigr\rangle_{A_{\bar{s}^j}^j},
\end{equation}
for $j\in[n]$. We note that $s^j\in\{0,1\}$ indicates which qubit in the pair $A^j$ is prepared in the computational basis $(\mathcal{D}_0)$ and which one in the Hadamard basis $(\mathcal{D}_1)$. For $i \in\{0,1\}$, the string $\bold{r}_i$ is prepared in the basis $\mathcal{D}_i$. $\mathcal{A}$ sends $\lvert \psi_{\bold{r}_0,\bold{r}_1}^{\bold{s}}\rangle$, i.e. the qubits $A^j_i$ with their labels $i,j$, to $\mathcal{B}$, for $i\in\{0,1\}$ and $j\in[n]$.

\item $\mathcal{B}$ chooses a random bit $c$, before receiving the qubits from $\mathcal{A}$. $\mathcal{B}$ measures $A^j_i$ in the basis $\mathcal{D}_{c}$, obtaining the bit outcome $d^j_i$, for $i \in \{0,1\}$ and $j \in [n]$. The outcomes define $\bold{d}_i=(d^1_i,d^2_i,\ldots,d^{n}_i)$, for $i\in\{0,1\}$. For $i\in\{0,1\}$, $\mathcal{B}$ transmits $c$, $\bold{d}_0$ and $\bold{d}_1$ to $\mathcal{B}_i$, who receives these in the causal past of $Q_i$.

\end{enumerate}

\begin{center}
\emph{Stage II}
\end{center}

\begin{enumerate}
\setcounter{enumi}{3}
\item $\mathcal{B}$ obtains his input $b\in\{0,1\}$ and gives the bit $b'=c\oplus b$ to $\mathcal{A}$.

\item For $i\in\{0,1\}$, $\mathcal{B}$ transmits $b$ to $\mathcal{B}_i$, with the transmission being completed in the causal past of $Q_i$.

\item For $i\in\{0,1\}$, $\mathcal{A}$ transmits $b'$ to $\mathcal{A}_i$, with the transmission being completed in the causal past of $Q_i$.

\item For $i \in \{0,1\}$, $\mathcal{A}_i$ obtains $\bold{x}_i$ in the causal past of $Q_i$, and transfers $\bold{t}_i=\bold{r}_{i\oplus b'}\oplus \bold{x}_i$ and $\bold{s}$ to $\mathcal{B}_i$ at $Q_i$.

\item Within $R_b$, $\mathcal{B}_b$ uses $\bold{s}$, $\bold{d}_0$, $\bold{d}_1$ and $c$ to compute $(d_{s^1\oplus c}^1,\ldots,d_{s^{n}\oplus c}^{n})$, which equals $\bold{r}_c$. Then, within $R_b$, $\mathcal{B}_b$ outputs $\bold{x}_{b}'=\bold{r}_c\oplus\bold{t}_b$, which equals $\bold{x}_b$.
\end{enumerate}



\subsection{Generalizing the SCOT protocols of Refs. \cite{PG15.1,PGK18.2} to the one-out-of-$m$ setting}

As described above, the SCOT protocols of Refs. \cite{PG15.1,PGK18.2} consider the one-out-of-two setting in which Alice has two input messages $\bold{x}_0$ and $\bold{x}_1$ and there are two spacelike separated output spacetime regions $R_0$ and $R_1$. In section \ref{section3} we generalize the definition of SCOT to the one-out-of-$m$ setting, where Alice has $m\geq 2$ input messages $\bold{x}_0,\bold{x}_1,\ldots,\bold{x}_{m-1}$, there are $m$ pair-wise spacelike separated output spacetime regions $R_0,R_1,\ldots,R_{m-1}$, Bob should obtain $\bold{x}_b$ in $R_b$ for his chosen $b\in\{0,1,\ldots,m-1\}$, Alice should not learn $b$, and Bob should not get $\bold{x}_i$ in $R_i$ for more than one $i$ from the set $\{0,1,\ldots,m-1\}$.

As discussed above, in the SCOT protocol of Ref. \cite{PG15.1}, Alice encodes a message $\bold{r}$ in a quantum state $\lvert \psi_{\bold{r}}^{\bold{s}}\rangle$, which further encodes the messages $\bold{x}_0$ and $\bold{x}_1$. In this protocol, Bob transmits the state $\lvert \psi_{\bold{r}}^{\bold{s}}\rangle$ received from Alice to his agent $\mathcal{B}_b$ having access to $R_b$, who is then able to decode $\bold{r}$, after receiving the basis label $\bold{s}$, and then uses $\bold{r}$ and $\bold{r}\oplus \bold{x}_b$ to decode $\bold{x}_b$ in $R_b$. As we detail in section \ref{section4}, this protocol can be straightforwardly generalized to the one-out-of-$m$ setting because the message $\bold{r}$ of the quantum state $\lvert \psi_{\bold{r}}^{\bold{s}}\rangle$ can be used to encode the messages $\bold{x}_i$ with the messages $\bold{x}_i\oplus \bold{r}$, for $i\in\{0,1,\ldots,m-1\}$.

On the other hand, generalizing the SCOT protocol of Ref. \cite{PGK18.2} to the one-out-of-$m$ setting is more complicated and interesting. We note from the discussion above, that the SCOT protocol of Ref. \cite{PGK18.2} works in two stages. Broadly, in the first stage, two messages $\bold{r}_0$ and $\bold{r}_1$ are encoded in a quantum state $\lvert \psi_{\bold{r}_0,\bold{r}_1}^{\bold{s}}\rangle$; and, in the second stage, these messages are used to encode further messages $\bold{x}_0$ and $\bold{x}_1$. Generalizing the first stage of this protocol to the one-out-of-$m$ setting requires to find a set of quantum states $\lvert \psi_{\bold{r}_0,\bold{r}_1,\ldots,\bold{r}_{m-1}}^{\bold{s}}\rangle$ that encodes $m$ messages $\bold{r}_0,\bold{r}_1,\ldots,\bold{r}_{m-1}$, while satisfying some security conditions. We have identified this stage with a task that we denote as one-out-of-$m$ distributed quantum access with classical memory, which we describe in section \ref{section5}, with a specific class of secure protocols for this task given in section \ref{section6}. Generalizing the second stage of the protocol to the one-out-of-$m$ setting is straightforward, and is done explicitly in section \ref{section7}.

Finally, further generalizations are discussed in section \ref{section8}. For example, we consider the case that the number $M\geq 2$ of output spacetime regions can be different to the number $m\geq2$ of Alice's inputs, and we discuss protocols for $k$-out-of-$m$ SCOT.

\section{One-out-of-$m$ spacetime-constrained oblivious transfer}
\label{section3}

\begin{figure}
\includegraphics{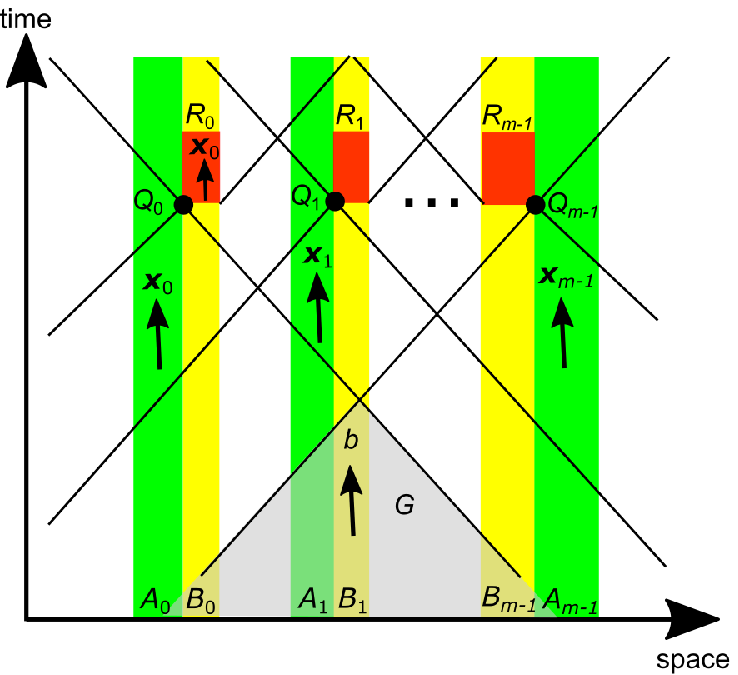}
 \caption{\label{fig2} Illustration of a setting for one-out-of-$m$ SCOT in a two-dimensional spacetime diagram, in a frame $\mathcal{F}$, of Minkowski spacetime. The world lines of the laboratories of Alice's agents $\mathcal{A}_0,\mathcal{A}_1,\ldots,\mathcal{A}_{m-1}$ (green rectangles), and of the laboratories of Bob's agents $\mathcal{B}_0,\mathcal{B}_1,\ldots,\mathcal{B}_{m-1}$ (yellow rectangles) are indicated. The small dots represent the spacetime points $Q_0,Q_1,\ldots,Q_{m-1}$. The thin diagonal lines represent light rays. The spacetime region $G$, consisting in the intersection of the causal pasts of $Q_0,Q_1,\ldots,Q_{m-1}$, is represented by the grey shaded area. The spacetime regions $R_i$, where Bob's agents must obtain Alice's inputs $\bold{x}_i$, correspond to the small red rectangles, for $i\in\{0,1,\ldots,m-1\}$. We note that $R_0,R_1,\ldots,R_{m-1}$ are pairwise spacelike separated. Alice inputs a message $\bold{x}_i$ in the causal past of $Q_i$, for $i\in\{0,1,\ldots,m-1\}$. Bob inputs a number $b\in\{0,1,\ldots,m-1\}$ in $G$ and outputs $\bold{x}_b$ in $R_b$. The case $b=0$ is illustrated.}
\end{figure}

We introduce a generalization of the definition of SCOT of Ref. \cite{PG15.1} to a one-out-of-$m$ setting, for any integer $m\geq 2$. Alice (Bob) has trusted agents who can process and communicate classical or quantum information at various locations in spacetime. But, Alice's agents do not trust Bob's agents, and vice versa. We assume that spacetime is Minkowski, or very close to Minkowski, as near the Earth's surface. Alice and Bob agree on a reference frame $\mathcal{F}$ in spacetime, and on $m$ pairwise spacelike separated \emph{output spacetime regions} $R_0,R_1,\ldots,R_{m-1}$; they also agree on a spacetime point $Q_i$ of $R_i$, for $i\in\text{I}_m=\{0,1,\ldots,m-1\}$. For $i\in\text{I}_m$, Alice inputs a message $\bold{x}_i$ in the causal past of $Q_i$, chosen from a set of possible messages previously agreed with Bob. Bob inputs a number $b\in\text{I}_m$ in the spacetime region $G$, which is defined as the intersection of the causal pasts of the spacetime points $Q_0,Q_1,\ldots,Q_{m-1}$, and he outputs $\bold{x}_b$ in $R_b$. We illustrate an example of a setting for one-out-of-$m$ SCOT in Fig. \ref{fig2}.

A one-out-of-$m$ SCOT protocol must satisfy correctness and security properties. Broadly speaking, the correctness property states that Bob obtains $\bold{x}_b$ in $R_b$, according to his input $b$, if Alice and Bob follow the protocol honestly. The security properties state: that if Bob follows the protocol honestly and Alice follows any dishonest strategy, Alice cannot learn Bob's input $b$ anywhere in spacetime; and that if Alice follows the protocol honestly and Bob follows any dishonest strategy, Bob cannot obtain $\bold{x}_i$ in $R_i$ and $\bold{x}_j$ in $R_j$, for any pair of different numbers $i$ and $j$ from the set $\text{I}_m$. We state these properties more precisely below in the ideal case in which Bob's outputs do not have any errors, and then in a scenario in which Bob's outputs have a small fraction of errors. We consider that $b$ is initially completely unknown to Alice, i.e. from her perspective, Bob chooses $b$ randomly from $\text{I}_m$. Similarly, we consider that $\bold{x}_i$  is initially completely unknown to Bob, i.e. from his perspective, Alice chooses $\bold{x}_i$ randomly from the previously agreed set, for $i\in\text{I}_m$.

\subsection{The ideal case of no errors}

\subsubsection{Correctness}

For $\epsilon_{\text{cor}}\geq 0$, we say a SCOT protocol is $\epsilon_{\text{cor}}-$correct if, when Alice and Bob follow the protocol honestly, the probability $P$ that Bob outputs $\bold{x}_b$ in $R_b$ satisfies $P\geq 1-\epsilon_{\text{cor}}$, for any inputs $\bold{x}_0,\bold{x}_1,\ldots,\bold{x}_{m-1}$ by Alice from the agreed sets, and for any input $b\in\text{I}_m$ by Bob. We say a SCOT protocol is perfectly correct if it is $0-$correct.

\subsubsection{Security}

For $\epsilon_{\text{Alice}}\geq 0$, we say a SCOT protocol is $\epsilon_{\text{Alice}}-$secure against dishonest Alice if, when Bob follows the protocol honestly and Alice follows any cheating strategy, the probability $P_{\text{Alice}}$ that Alice guesses Bob's input $b$ anywhere in spacetime satisfies $P_{\text{Alice}}\leq \frac{1}{m}+\epsilon_{\text{Alice}}$. We say a SCOT protocol is perfectly secure against dishonest Alice if it is $0-$secure against dishonest Alice. We say a SCOT protocol is unconditionally secure against dishonest Alice if it is $\epsilon_{\text{Alice}}-$secure against dishonest Alice with $\epsilon_{\text{Alice}}$ approaching zero by increasing some security parameter, for any cheating strategy by Alice that is allowed by quantum theory and relativity.

For $\epsilon_{\text{Bob}}\geq 0$, we say a SCOT protocol is $\epsilon_{\text{Bob}}-$secure against dishonest Bob if, when Alice follows the protocol honestly and Bob follows any cheating strategy, the probability $P_{\text{Bob}}$ that Bob outputs $\bold{x}_i$ in $R_i$ and $\bold{x}_j$ in $R_j$ satisfies $P_{\text{Bob}}\leq \epsilon_{\text{Bob}}$,  for any pair of different numbers $i,j$ from the set $\text{I}_m$. Ideally, we would define a SCOT protocol to be unconditionally secure against dishonest Bob if it is $\epsilon_{\text{Bob}}-$secure against dishonest Bob with $\epsilon_{\text{Bob}}$ approaching $\frac{1}{d}$ by increasing some security parameter, for any cheating strategy by Bob that is allowed by quantum theory and relativity, where $d$ is the minimum of the number of possible values of $\bold{x}_i$, for $i\in\text{I}_m$. That is, ideally, we would like to guarantee that when a security parameter tends to infinity, Bob should not be able to do better than following the honest protocol to obtain some $\bold{x}_b$ in $R_b$ and to make a random guess of some other $\bold{x}_j$ in $R_j$. However, here we can satisfy a weaker definition of security: we say a SCOT protocol is unconditionally secure against dishonest Bob if it is $\epsilon_{\text{Bob}}-$secure against dishonest Bob with $\epsilon_{\text{Bob}}$ approaching zero by increasing the size of Alice's input messages and possibly by increasing some other security parameters, for any cheating strategy by Bob that is allowed by quantum theory and relativity.

\subsection{Tolerating a small fraction of errors}

We generalize the previous definition of one-out-of-$m$ SCOT to allow a small fraction of errors in Bob's output. We consider that Alice and Bob agree that Alice's inputs are of the form $\bold{x}_i\in\{0,1\}^{n_i}$, for an agreed number $n_i\in\mathbb{N}$, and for $i\in\text{I}_m$. Alice and Bob agree on parameters $\gamma_i\geq 0$, for $i\in\text{I}_m$.  In following sections we present protocols to implement one-out-of-$m$ SCOT considering the particular case $n_i=n$ and $\gamma_i=\gamma$, for $i\in\text{I}_m$.

\subsubsection{Correctness}

For $\epsilon_{\text{cor}}\geq 0$, we say a SCOT protocol is $\epsilon_{\text{cor}}-$correct if, when Alice and Bob follow the protocol honestly, the probability $P$ that Bob outputs a message $\bold{x}_b'$ in $R_b$ satisfying $d(\bold{x}_b',\bold{x}_b)\leq n_b\gamma_b$ satisfies $P\geq 1-\epsilon_{\text{cor}}$, for any inputs $\bold{x}_0,\bold{x}_1,\ldots,\bold{x}_{m-1}$ by Alice from the agreed sets, and for any input $b\in\text{I}_m$ by Bob. We say a SCOT protocol is perfectly correct if it is $0-$correct.

\subsubsection{Security}

For $\epsilon_{\text{Bob}}\geq 0$, we say a SCOT protocol is $\epsilon_{\text{Bob}}-$secure against dishonest Bob if, when Alice follows the protocol honestly and Bob follows any cheating strategy, the probability $P_{\text{Bob}}$ that Bob outputs messages $\bold{x}_i'$ in $R_i$ and $\bold{x}_j'$ in $R_j$ satisfying $d(\bold{x}_i',\bold{x}_i)\leq n_i\gamma_i$ and $d(\bold{x}_j',\bold{x}_j)\leq n_j\gamma_j$ satisfies $P_{\text{Bob}}\leq \epsilon_{\text{Bob}}$, for any pair of different numbers $i$ and $j$ from the set $\text{I}_m$. We say a SCOT protocol is unconditionally secure against dishonest Bob if it is $\epsilon_{\text{Bob}}-$secure against dishonest Bob with $\epsilon_{\text{Bob}}$ approaching zero by increasing the size of Alice's input messages and possibly some other security parameters, for any cheating strategy by Bob that is allowed by quantum theory and relativity. Security against dishonest Alice is defined as in the ideal case of no errors.


\section{An unconditionally secure one-out-of-$m$ SCOT protocol $\mathcal{P}_{\text{QC}}$ with long-distance quantum communication}
\label{section4}

We introduce the protocol $\mathcal{P}_{\text{QC}}$ for one-out-of-$m$ SCOT, which extends straightforwardly the protocol for one-out-of-two SCOT of Ref. \cite{PG15.1}. The label `QC' stands for `quantum communication', as the protocol $\mathcal{P}_{\text{QC}}$ requires quantum communication among Bob's distant agents. The protocol $\mathcal{P}_{\text{QC}}$ uses a subroutine $\mathcal{P}_{\text{SR}}$ introduced below.

The setting is the following. Alice has trusted agents $\mathcal{A}$ and $\mathcal{A}_i$, for $i\in\text{I}_m$. Bob has trusted agents $\mathcal{B}$ and $\mathcal{B}_i$, for $i\in\text{I}_m$. Alice and Bob agree in a reference frame $\mathcal{F}$ and in $m$ pairwise spacelike separated output spacetime regions $R_0, R_1,\ldots,R_{m-1}$. Each of Alice's (Bob's) agents controls a secure laboratory. It is useful to consider that the agents $\mathcal{A}$ and $\mathcal{B}$ have adjacent laboratories, and that the agents $\mathcal{A}_i$ and $\mathcal{B}_i$ have adjacent laboratories, for $i\in\text{I}_m$. Alice's agents share secure and authenticated classical channels, and Bob's agents share secure and authenticated quantum channels. There is a quantum channel between $\mathcal{A}$ and $\mathcal{B}$. There is a classical channel between $\mathcal{A}_i$ and $\mathcal{B}_i$, for $i\in\text{I}_m$.

\subsection{The subroutine $\mathcal{P}_{\text{SR}}$}

The following protocol was used as a subroutine in Ref. \cite{PG15.1} for the case $m=2$. We extend this protocol here for the case $m\geq2$ and denote it as $\mathcal{P}_{\text{SR}}$.

\begin{enumerate}

\item Alice's agent $\mathcal{A}$ encodes a random $n-$bit string $\bold{r}$ in a quantum state $\lvert \psi_{\bold{r}}^{\bold{s}}\rangle=\bigotimes_{j\in[n]}\lvert \psi_{r^j}^{s^j}\rangle$ of $n$ BB84 states, where $\bold{s}$ is a random $n-$bit string denoting the bases. $\mathcal{A}$ sends $\lvert \psi_{\bold{r}}^{\bold{s}}\rangle$ to $\mathcal{B}$, who receives it in $G$. 

\item $\mathcal{B}$ obtains his input number $b\in\text{I}_m$ in $G$ and redirects the received state $\lvert \psi_{\bold{r}}^{\bold{s}}\rangle$ to his colleague $\mathcal{B}_b$, who receives it in the causal past of at least one spacetime point of $R_b$.

\item For $i\in\text{I}_m$, $\mathcal{A}$ sends a copy of $\bold{s}$ to her colleague $\mathcal{A}_i$, who receives it in the causal past of $Q_i$. 

\item For $i\in\text{I}_m$, $\mathcal{A}_i$ gives $\bold{s}$ to $\mathcal{B}_i$ at the spacetime point $Q_i$.

\item $\mathcal{B}_b$ measures the quantum state $\lvert \psi_{\bold{r}}^{\bold{s}}\rangle$ in the basis labeled by $\bold{s}$, and obtains a $n-$bit string $\bold{r}'$ in $R_b$.

\end{enumerate}

\subsubsection{Correctness}

In the ideal case in which there are not any errors nor any losses, Bob's output $\bold{r}'$ equals $\bold{r}$ with unit probability.

\subsubsection{Security against dishonest Alice}

Since Bob does not transmit any physical systems to Alice, Alice cannot obtain any information about Bob's input $b$. Thus, Alice cannot guess Bob's input $b$ with probability greater than $\frac{1}{m}$.

\subsubsection{Security against dishonest Bob}

For the case $m=2$, it was shown in Ref. \cite{PG15.1} that if Alice follows the protocol honestly and Bob follows an arbitrary cheating strategy allowed by quantum theory and relativity, the probability $P_{\text{Bob}}$ that Bob outputs $n-$bit strings $\bold{r}_0=\bold{r}$ in $R_0$ and $\bold{r}_1=\bold{r}$ in $R_1$ satisfies $P_\text{Bob}\leq \bigl(\frac{1}{2}+\frac{1}{2\sqrt{2}}\bigr)^n$. It was also shown that for a sufficiently small positive parameter $\gamma$, the probability $P_{\text{Bob}}^\gamma$ that Bob outputs $n-$bit strings $\bold{r}_0$ in $R_0$ and $\bold{r}_1$ in $R_1$ satisfying $d(\bold{r}_0,\bold{r})\leq n\gamma$ and $d(\bold{r}_1,\bold{r})\leq n\gamma$ decreases exponentially with $n$. In particular, we have $P_{\text{Bob}}^\gamma \leq (q_\gamma)^n$, where $q_\gamma=2^{2h(\gamma)}\bigl(\frac{1}{2}+\frac{1}{2\sqrt{2}}\bigr)<1$, for $\gamma< 0.015$, and where $h(\gamma)=-\gamma\log_2 \gamma -(1-\gamma)\log_2(1-\gamma)$ denotes the binary entropy of $\gamma$ \cite{PG15.1}. As we show below, these security properties hold for all $m\geq 2$.

Consider any pair of different numbers $i,j\in\text{I}_m$. In an arbitrary cheating strategy by Bob allowed by quantum theory and relativity in which he tries to output $\bold{r}_i=\bold{r}$ in $R_i$ and  $\bold{r}_j=\bold{r}$ in $R_j$, $\mathcal{B}$ applies some quantum operation on the quantum state received from Alice and outputs two quantum systems $B_0$ and $B_1$ that he sends to his colleagues $\mathcal{B}_i$ and $\mathcal{B}_j$, respectively. Then, after receiving $\bold{s}$ form Alice's agents, $\mathcal{B}_i$ and $\mathcal{B}_j$ apply respective measurements $\tilde{\text{M}}_{0,\bold{s}}$ and $\tilde{\text{M}}_{1,\bold{s}}$, and obtain respective measurement outcomes $\bold{r}_i$ and $\bold{r}_j$. This cheating strategy is the same as for the case $m=2$, where $i=0$ and $j=1$. Thus, we see that the security conditions stated in the previous paragraph hold for any pair of different numbers $i,j\in\text{I}_m$, both in the case of perfect outcomes, for which we have $P_\text{Bob}\leq \bigl(\frac{1}{2}+\frac{1}{2\sqrt{2}}\bigr)^n$; and in the case in which a small fraction of errors $\gamma\geq 0$ is tolerated, in particular $P_{\text{Bob}}^\gamma \leq (q_\gamma)^n$, where $q_\gamma=2^{2h(\gamma)}\bigl(\frac{1}{2}+\frac{1}{2\sqrt{2}}\bigr)<1$, for $\gamma<0.015$.

\subsection{The one-out-of-$m$ SCOT protocol $\mathcal{P}_{\text{QC}}$}

\begin{enumerate}

\item Alice and Bob implement the subroutine $\mathcal{P}_{\text{SR}}$, where $\bold{r}\in\{0,1\}^n$ is the message encoded by $\mathcal{A}$ in $G$ and $b\in\text{I}_m$ is $\mathcal{B}$'s input in $G$, and where $\bold{r}'\in\{0,1\}^n$ is $\mathcal{B}_b$'s output in $R_b$.

\item For $i\in\text{I}_m$, $\mathcal{A}$ sends a copy of $\bold{r}$ to $\mathcal{A}_i$, who receives it in the causal past of $Q_i$.

\item For $i\in\text{I}_m$, $\mathcal{A}_i$ obtains her input message $\bold{x}_i\in\{0,1\}^n$ in the causal past of $Q_i$.

\item For $i\in\text{I}_m$, $\mathcal{A}_i$ gives $\bold{t}_i=\bold{x}_i\oplus \bold{r}$ to $\mathcal{B}_i$ at the spacetime point $Q_i$.

\item $\mathcal{B}_b$ computes the message $\bold{x}_b'=\bold{r}'\oplus \bold{t}_b$ and outputs it in $R_b$.

\end{enumerate}

\subsubsection{Comments}

Different variations of this protocol can be considered. For example, Bob's agents having quantum memories have more freedom on the time at which they receive, process and transmit classical and quantum information. On the other hand, if Bob does not have any quantum memories, Bob's agent $\mathcal{B}$ must redirect the  quantum states as soon as he receives them from Alice's agent $\mathcal{A}$; and the transmission of the quantum state from $\mathcal{A}$ to $\mathcal{B}$ must be completed within a sufficiently short time interval so that Bob's agent $\mathcal{B}_b$ is able to complete the corresponding quantum measurement on it within the output spacetime region $R_b$; a physical implementation of this protocol seems plausible in some scenarios using photons as the physical systems encoding the quantum states, for example. Additionally, here we have considered that the subroutine $\mathcal{P}_{\text{SR}}$ is implemented with BB84 states, but generalizations with other sets of non-mutually orthogonal states can be devised.
Furthermore, we note that Bob's secure and authenticated quantum channels can be implemented via the teleportation \cite{teleportation} protocol if Bob's agents share entangled states and authenticated classical channels, or via the quantum one-time pad \cite{AMTW00} if Bob's agents share secret classical keys and authenticated quantum channels.

\subsubsection{Correctness}

In the ideal case that there are not any errors nor any losses, Bob outputs $\bold{r}'=\bold{r}$ in $R_b$ in the subroutine $\mathcal{P}_{\text{SR}}$, hence, Bob outputs $\bold{x}_b'=\bold{x}_b$ in $R_b$ in the protocol $\mathcal{P}_{\text{QC}}$. In an implementation of $\mathcal{P}_{\text{SR}}$ in which Bob outputs $\bold{r}'$ in $R_b$ satisfying $d(\bold{r}',\bold{r})\leq \gamma n$, Bob's output $\bold{x}_b'$ in $R_b$ satisfies $d(\bold{x}_b',\bold{x}_b)\leq \gamma n$, for $\gamma\geq0$.

\subsubsection{Security against dishonest Alice}

Like in the subroutine $\mathcal{P}_{\text{SR}}$, Bob does not transmit any physical systems to Alice in the protocol $\mathcal{P}_{\text{QC}}$, hence, Alice cannot obtain any information about Bob's input $b$. Thus, Alice cannot guess Bob's input $b$ with probability greater than $\frac{1}{m}$. Therefore the protocol $\mathcal{P}_{\text{QC}}$ is perfectly secure against dishonest Alice.

\subsubsection{Security against dishonest Bob}

For any pair of different numbers $i,j\in\text{I}_m$, the probability $P_\text{Bob}$ that Bob outputs $\bold{x}_i'\in\{0,1\}^n$ in $R_i$ and $\bold{x}_j'\in\{0,1\}^n$ in $R_j$ in the protocol $\mathcal{P}_{\text{QC}}$ satisfying $d(\bold{x}_i',\bold{x}_i)\leq n\gamma$ and $d(\bold{x}_j',\bold{x}_j)\leq n\gamma$ is equal to the probability that Bob outputs $\bold{r}_i\in\{0,1\}^n$ in $R_i$ and $\bold{r}_j\in\{0,1\}^n$ in $R_j$ in the subroutine $\mathcal{P}_{\text{SR}}$ satisfying $d(\bold{r}_i,\bold{r})\leq n\gamma$ and $d(\bold{r}_j,\bold{r})\leq n\gamma$. Thus, from the security properties of the subroutine $\mathcal{P}_{\text{SR}}$, it follows straightforwardly that if errors are not tolerated, i.e. if $\gamma=0$, the protocol $\mathcal{P}_{\text{QC}}$ is
$\epsilon-$secure against dishonest Bob, with $\epsilon=\bigl(\frac{1}{2}+\frac{1}{2\sqrt{2}}\bigr)^n$. Since $\epsilon$ decreases exponentially with $n$, $\mathcal{P}_{\text{QC}}$ is unconditionally secure against dishonest Bob. 

Similarly, if the protocol $\mathcal{P}_{\text{QC}}$ tolerates a fraction of errors $\gamma$, the protocol $\mathcal{P}_{\text{QC}}$ is $\epsilon_\gamma-$secure against dishonest Bob, with $\epsilon_\gamma=(q_\gamma)^n$, where $q_\gamma=2^{2h(\gamma)}\bigl(\frac{1}{2}+\frac{1}{2\sqrt{2}}\bigr)<1$, for $\gamma<0.015$. Since $\epsilon_\gamma$ decreases exponentially with $n$, for $\gamma<0.015$, $\mathcal{P}_{\text{QC}}$ is unconditionally secure against dishonest Bob in this case too.

\section{One-out-of-$m$ distributed quantum access with classical memory}
\label{section5}

In this section we introduce a task that we denote as one-out-of-$m$ distributed quantum access with classical memory (DQACM), which for simplicity of the exposition we often refer to simply as DQACM. This task works as a primitive to construct one-out-of-$m$ SCOT protocols in which Bob's agents only need to communicate classical information. 

We define one-out-of-$m$ DQACM as follows. Previous to implementing a DQACM protocol, Alice and Bob agree on the integer $m\geq2$; on finite sets of classical messages $\Omega_0,\Omega_1,\ldots, \Omega_{m-1}$, $\Omega_{\text{outcome}}\subseteq \Omega_0\times\Omega_1\times\cdots\times\Omega_{m-1}$, and $\Lambda_{\text{basis}}$; and on a set of quantum states $\Delta=\{\bigl\lvert \Psi_{\bold{r}}^{\bold{s}}\bigr\rangle\big\vert\bold{r}\in \Omega_{\text{outcome}},\bold{s}\in\Lambda_{\text{basis}}\}$. In general, we may set $\Omega_{\text{outcome}}\subseteq \Omega_0\times\Omega_1\times\cdots\times\Omega_{m-1}$, but here we consider $\Omega_{\text{outcome}}= \Omega_0\times\Omega_1\times\cdots\times\Omega_{m-1}$. A DQACM protocol consists of two stages.

In stage I, Alice encodes a string of messages $\bold{r}=(\bold{r}_0,\bold{r}_1,\ldots, \bold{r}_{m-1})\in \Omega_{\text{outcome}}$ in a quantum state $\bigl\lvert \Psi_{\bold{r}}^{\bold{s}}\bigr\rangle_A\in\Delta$ of a quantum system $A$, using the extra classical message $\bold{s}\in\Lambda_{\text{basis}}$. The message $\bold{s}$ may indicate, for example, the basis used by Alice to prepare the state $\bigl\lvert \Psi_{\bold{r}}^{\bold{s}}\bigr\rangle$, from a set of possibly non-mutually orthogonal bases. Alice gives the quantum state $\bigl\lvert \Psi_{\bold{r}}^{\bold{s}}\bigr\rangle_A$ to Bob. We consider that Alice's inputs are initially secret to Bob, i.e. from Bob's perspective, Alice chooses $\bold{r}_i$ randomly from $\Omega_i$, for $i\in\text{I}_m$, and $\bold{s}$ randomly from $\Lambda_{\text{basis}}$. Bob inputs a number $c\in\text{I}_m$, initially secret to Alice, i.e. from Alice's perspective, Bob chooses $c$ randomly from $\text{I}_m$. Bob applies a quantum measurement $\text{M}_c$ on the received quantum state $\bigl\lvert \Psi_{\bold{r}}^{\bold{s}}\bigr\rangle_A$ and obtains a classical measurement outcome $\bold{d}$. 

Stage II consists of two steps. In the first step, Alice gives the classical message $\bold{s}$ to Bob. In the second step, Bob applies a function $f$ on $(c,\bold{d},\bold{s})$ giving as output a classical message $\bold{r}_c'$, i.e. $\bold{r}_c'=f(c,\bold{d},\bold{s})$. 

\begin{figure}
\includegraphics{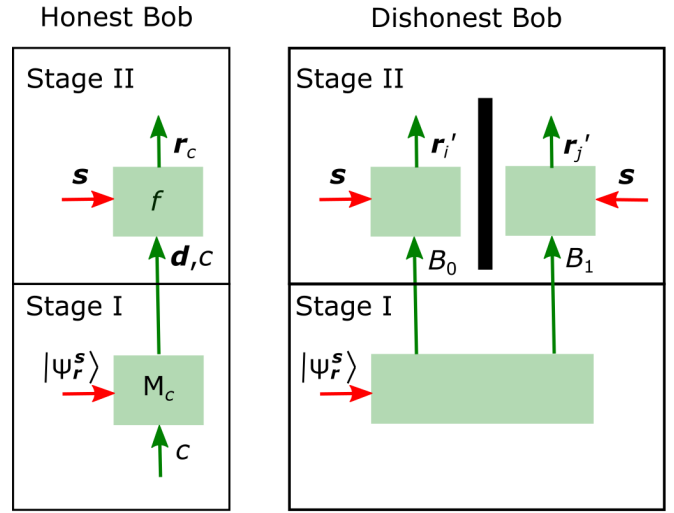}
 \caption{\label{fig3} One-out-of-$m$ DQACM implemented between Alice and Bob. The horizontal red arrows represent communication from Alice to Bob. The vertical green arrows represent inputs and outputs by Bob. The green boxes represent processing by Bob. The task consists in two stages. By definition, we consider that Alice implements the task honestly. Alice's actions consist, in stage I, in giving to Bob a quantum state $\lvert\Psi_{\bold{r}}^{\bold{s}}\rangle$ encoding the string $\bold{r}=(\bold{r}_0,\bold{r}_1,\ldots,\bold{r}_{m-1})$ in a basis labeled by a classical string $\bold{s}$; and in stage II, in giving $\bold{s}$ to Bob. Left: Bob implements the task honestly. In stage I, Bob inputs a number $c\in\{0,1,\ldots,m-1\}$ and applies a quantum measurement $\text{M}_c$, depending on $c$, on the received quantum state $\lvert\Psi_{\bold{r}}^{\bold{s}}\rangle$, and obtains a classical measurement outcome $\bold{d}$. In stage II, Bob applies a function $f$ on $c$, $\bold{d}$ and $\bold{s}$, and obtains Alice's input $\bold{r}_c$ (or a message $\bold{r}_c'$ close to $\bold{r}_c$ according to a predetermined threshold). Right: Bob follows a dishonest strategy. In stage I, Bob applies an arbitrary quantum operation (large green box) on the received quantum state $\lvert\Psi_{\bold{r}}^{\bold{s}}\rangle$ and outputs two quantum systems $B_0$ and $B_1$, which he sends to his agents $\mathcal{B}_0$ and $\mathcal{B}_1$. In stage II, Bob's agents $\mathcal{B}_0$ and $\mathcal{B}_1$ (small green boxes) cannot communicate, as represented by the thick black line. Bob's agent $\mathcal{B}_0$ ($\mathcal{B}_1$) use the classical message $\bold{s}$ and the received quantum system $B_0$ ($B_1$) to output a classical message $\bold{r}_i'$ ($\bold{r}_j'$). In a secure one-out-of-$m$ DQACM protocol, the probability that Bob's inputs $\bold{r}_i'$ and $\bold{r}_j'$ are equal to Alice's inputs $\bold{r}_i$ and $\bold{r}_j$  (or close to Alice's inputs $\bold{r}_i$ and $\bold{r}_j$ according to a predetermined threshold) is negligible, for any pair of different numbers $i,j\in\{0,1,\ldots,m-1\}$.}
\end{figure}

Ideally, a protocol for one-out-of-$m$ DQACM should satisfy a correctness and a security condition. Broadly speaking, the correctness property says that if Alice and Bob follow the protocol honestly, Bob's output $\bold{r}_c'$ is equal to Alice's input $\bold{r}_c$, or is sufficiently close to Alice's input $\bold{r}_c$, according to a predetermined threshold. The security condition states that in distributed cheating strategies by Bob involving two agents of Bob, $\mathcal{B}_0$ and $\mathcal{B}_1$, who receive Alice's message $\bold{s}$, and who cannot communicate with each other after receiving $\bold{s}$, $\mathcal{B}_0$ and $\mathcal{B}_1$ cannot both output Alice's inputs $\bold{r}_i$ and $\bold{r}_j$ -- or messages sufficiently close to $\bold{r}_i$ and $\bold{r}_j$, according to a predetermined threshold -- respectively, for any pair of different numbers $i,j\in\text{I}_m$. Fig. \ref{fig3} illustrates the task of one-out-of-$m$ DQACM.
 
We see that by construction, in the defined task of one-out-of-$m$ DQACM, Alice remains oblivious to Bob's input $c$, as she does not receive any physical system from Bob. This \emph{obliviousness} property, as well as the properties of correctness and security against dishonest Bob allow us to use one-out-of-$m$ DQACM to construct correct and secure one-out-of-$m$ SCOT.

We explain why we have denoted this task as `one-out-of-$m$ distributed quantum access with classical memory'. First, in the honest protocol Bob chooses to \emph{access one out of $m$} messages that Alice encodes in a \emph{quantum} state. Second, in the honest protocol, we can consider that Bob has only \emph{classical memory}, as he receives the quantum state from Alice and is forced to apply a quantum measurement and then apply further classical processing on his outcomes after receiving the classical message $\bold{s}$ from Alice. Third, in a dishonest cheating strategy by Bob, we can consider that he outputs two quantum systems $B_0$ and $B_1$ and \emph{distributes} them to his agents $\mathcal{B}_0$ and $\mathcal{B}_1$, who without communicating -- because they must obtain their outputs at spacelike separated spacetime regions, for instance -- must respectively output Alice's inputs $\bold{r}_i$ and $\bold{r}_j$ after receiving Alice's classical message $\bold{s}$, for some pair of different numbers $i,j\in\text{I}_m$. 

We define precisely the correctness and security conditions in two broad scenarios: an ideal scenario in which there are not errors in Bob's output, and a more general scenario in which there is a small fraction of errors in Bob's output.


\subsection{The ideal case of no errors}

\subsubsection{Correctness}
For $\epsilon_{\text{cor}}\geq 0$, we say a protocol to implement one-out-of-$m$ DQACM is $\epsilon_{\text{cor}}-$correct if, when Alice and Bob follow the protocol honestly, the probability $P$ that Bob outputs $\bold{r}_c$ satisfies $P\geq 1-\epsilon_{\text{cor}}$, for any input $c\in\text{I}_m$ by Bob. We say a  protocol to implement one-out-of-$m$ DQACM is perfectly correct if it is $0-$correct.

\subsubsection{Security}

For $\epsilon_{\text{Bob}}\geq 0$, we say a protocol to implement one-out-of-$m$ DQACM is $\epsilon_{\text{Bob}}-$secure against dishonest Bob if, when Alice follows the protocol honestly, for any pair of different numbers $i$ and $j$ from the set $\text{I}_m$, for any quantum operation $O$ independent of $\bold{r}$ and independent of $\bold{s}$ applied by Bob on the quantum state $\lvert \Psi_ {\bold{r}}^{\bold{s}}\rangle_A$ that produces at least two quantum systems $B_0$ and $B_1$, and for any sets of quantum measurements $\bigl\{\tilde{\text{M}}_0^\bold{s}\bigr\}_{\bold{s}\in\Lambda_{\text{basis}}}$ and $\bigl\{\tilde{\text{M}}_1^\bold{s}\bigr\}_{\bold{s}\in\Lambda_{\text{basis}}}$, the probability to obtain respective outcomes $\bold{r}_i$ and $\bold{r}_j$, by respectively applying $\tilde{\text{M}}_0^\bold{s}$ on $B_0$ and $\tilde{\text{M}}_1^\bold{s}$ on $B_1$, is not greater than $\epsilon_{\text{Bob}}$. We say a protocol to implement one-out-of-$m$ DQACM is unconditionally secure against dishonest Bob if it is $\epsilon_{\text{Bob}}-$secure against dishonest Bob with $\epsilon_{\text{Bob}}$ approaching zero by increasing the size of Alice's input messages and possibly some other security parameters.

\subsection{Tolerating a small fraction of errors}

We generalize the previous definition of one-out-of-$m$ DQACM to allow a small fraction of errors in Bob's output. We consider that Alice and Bob agree that Alice's inputs are of the form $\bold{r}_i\in\{0,1\}^{n_i}$, for an agreed number $n_i\in\mathbb{N}$, and for $i\in\text{I}_m$. Alice and Bob agree on parameters $\gamma_i\geq 0$, for $i\in\text{I}_m$. 


\subsubsection{Correctness}

For $\epsilon_{\text{cor}}\geq 0$, we say a protocol to implement one-out-of-$m$ DQACM is $\epsilon_{\text{cor}}-$correct if, when Alice and Bob follow the protocol honestly, the probability $P$ that Bob outputs a message $\bold{r}_c'$ satisfying $d(\bold{r}_c',\bold{r}_c)\leq n_c\gamma_c$ satisfies $P\geq 1-\epsilon_{\text{cor}}$, for any input $c\in\text{I}_m$ by Bob. We say a protocol to implement one-out-of-$m$ DQACM is perfectly correct if it is $0-$correct.

\subsubsection{Security}

For $\epsilon_{\text{Bob}}\geq 0$, we say a protocol to implement one-out-of-$m$ DQACM is $\epsilon_{\text{Bob}}-$secure against dishonest Bob if, when Alice follows the protocol honestly, for any pair of different numbers $i,j\in \text{I}_m$, for any quantum operation $O$ independent of $\bold{r}$ and independent of $\bold{s}$ applied by Bob on the quantum state $\lvert \Psi_ {\bold{r}}^{\bold{s}}\rangle_A$ that produces at least two quantum systems $B_0$ and $B_1$, and for any sets of quantum measurements $\bigl\{\tilde{\text{M}}_0^\bold{s}\bigr\}_{\bold{s}\in\Lambda_{\text{basis}}}$ and $\bigl\{\tilde{\text{M}}_1^\bold{s}\bigr\}_{\bold{s}\in\Lambda_{\text{basis}}}$, the probability to obtain respective outcomes $\bold{r}_i'$ and $\bold{r}_j'$ satisfying $d(\bold{r}_i',\bold{r}_i)\leq \gamma_i n_i$ and $d(\bold{r}_j',\bold{r}_j)\leq \gamma_j n_j$, by respectively applying $\tilde{\text{M}}_0^\bold{s}$ on $B_0$ and $\tilde{\text{M}}_1^\bold{s}$ on $B_1$, is not greater than $\epsilon_{\text{Bob}}$. We say a protocol to implement one-out-of-$m$ DQACM is unconditionally secure against dishonest Bob if it is $\epsilon_{\text{Bob}}-$secure against dishonest Bob with $\epsilon_{\text{Bob}}$ approaching zero by increasing the size of Alice's input messages and possibly some other security parameters.

\section{A class $\mathcal{C}$ of unconditionally secure protocols for one-out-of-$m$ DQACM}

\label{section6}
We introduce a class $\mathcal{C}$ of protocols to implement one-out-of-$m$ DQACM. We define a set $\Omega$ with $l$ distinct elements, for some integer $l\geq 2$. We define $\Lambda=\{(a_0,a_1,\ldots,a_{m-1})\vert a_i\in \text{I}_m ~\text{and}~a_i\neq a_{i'}~\text{if}~i\neq{i'},~\text{for}~i,i'\in\text{I}_m\}$, i.e. the set $\Lambda$ is in one-to-one correspondence with the set of permutations of $m$ distinct elements. We define $\Omega_{\text{outcome}}=\Omega^{nm}$ and $\Lambda_{\text{basis}}=\Lambda^n$. This means that we consider strings $\bold{s}=(s^1,s^2,\ldots,s^{n})\in\Lambda^n$, $\bold{r}_i=(r_i^1,r_i^2,\ldots,r_{i}^n)\in \Omega^n$ and $\bold{r}=(\bold{r}_0,\bold{r}_1,\ldots,\bold{r}_{m-1})\in\Omega^{nm}$, i.e. with $s^j=(s_0^j,s_1^j,\ldots,s^{j}_{m-1})\in\Lambda$ and $r_i^j\in\Omega$, for $j\in[n]$ and $i\in\text{I}_m$. We note that the number of elements of the set $\Lambda$ is $m!$, hence, the number of elements of the set $\Lambda^n$ is $(m!)^n$. Similarly, as the number of elements of the set $\Omega$ is $l$, the number of elements of the set $\Omega^n$ is $(l)^n$.

Alice generates the quantum state $\lvert \Psi_{\bold{r}}^{\bold{s}}\rangle_A$ encoded in a quantum system with Hilbert space $A$, where the strings  $\bold{s}\in\Lambda^n$ and $\bold{r}_i\in\Omega^n$ are randomly generated by her, for $i\in\text{I}_m$. Alice gives the quantum state $\lvert \Psi_{\bold{r}}^{\bold{s}}\rangle_A$ to Bob. The quantum state is of the form
\begin{equation}
\label{x1}
\bigl\lvert \Psi_{\bold{r}}^{\bold{s}}\bigr\rangle_A=\bigotimes_{\substack{
   i\in \text{I}_m \\  j\in[n]  }} \bigl\lvert \alpha_{r_i^j}^{i}\bigr\rangle_{A_{s_i^j}^{j}},
\end{equation}
where $\bold{r}=(\bold{r}_0,\bold{r}_1,\ldots,\bold{r}_{m-1})$ denotes the string of $m$ messages encoded by Alice, and where the Hilbert space $A$ is a tensor product of
$mn$ Hilbert spaces $A_i^j$, as follows,
\begin{equation}
\label{x2}
A=\bigotimes_{\substack{
   i\in \text{I}_m \\  j\in[n]  }} A_i^j,
\end{equation}
with the dimension of the Hilbert space $A_i^j$ being equal to $l$ for all $j\in[n]$ and all $i\in\text{I}_m$, and where $\mathcal{D}_i=\{\lvert \alpha_r^i\rangle\}_{r\in\Omega}$ is an orthonormal basis of an $l$-dimensional Hilbert space, for $i\in\text{I}_m$. We define
\begin{equation}
\label{x3}
\lambda=\max_{i\neq i'} \bigl\lvert \bigl\langle \alpha_r^i\vert \alpha_{r'}^{i'}\bigr\rangle\bigr\rvert^2,
\end{equation}
where the maximum is taken over all $r,r'\in \Omega$ and over all $i,i'\in\text{I}_m$ with $i\neq i'$.

Bob generates his input $c\in\text{I}_m$ and applies a quantum measurement $\text{M}_c$ on the received quantum state $\lvert \Psi_{\bold{r}}^{\bold{s}}\rangle_A$. The quantum measurement $\text{M}_c$ consists in measuring the quantum subsystem $A_i^j$ of $A$ in the basis $\mathcal{D}_c$, whose classical measurement outcome is denoted by $d_i^j$, for $j\in[n]$ and $i\in\text{I}_m$. We denote $d^j=(d_0^j,d_1^j,\ldots,d_{m-1}^j)$, for $j\in[n]$, and Bob's total classical measurement outcome by $\bold{d}=(d^1,d^2,\ldots,d^n)$. We note that the outcomes satisfy $d^j_{s^j_c}=r^j_c$, because the quantum system $A^j_{s^j_c}$ is prepared by Alice in the quantum state $\bigl\lvert \alpha_{r_c^j}^{c}\bigr\rangle$, i.e. in the basis $\mathcal{D}_c$ encoding the classical outcome $r_c^j$, for $j\in[n]$. Thus, for $c\in\text{I}_m$, following the protocol honestly and using $c$, $\bold{d}$ and $\bold{s}$, Bob can decode Alice's input $\bold{r}_c$, i.e. there exists a function $f$ that when applied on $(c,\bold{d},\bold{s})$ gives as output Alice's input $\bold{r}_c$. Therefore, in the ideal case that there are not errors nor losses, the protocols of this class are perfectly correct.


In order to guarantee security against dishonest Bob, the set of bases $\{\mathcal{D}_i\}_{i\in\text{I}_m}$ is chosen to satisfy the constraint
\begin{equation}
\label{x4}
\lambda<1. 
\end{equation}
For fixed values of $l$ and $m$, the smaller the value of $\lambda$ is, the greater the security that can be guaranteed. For this reason, it is preferable that the bases $\mathcal{D}_i$ are mutually unbiased, i.e. that $\lvert \langle \alpha_r^i\vert \alpha_{r'}^{i'}\rangle\rvert^2=l^{-1}$ for all $r,r'\in\Omega$ and all $i,i'\in\text{I}_m$ with $i\neq i'$, in which case $\lambda=l^{-1}$.
For example, in the case $m=2$, we can set $l=2$, and a pair of mutually unbiased bases can be given by the computational and Hadamard bases. 

In order to quantitatively prove the security against dishonest Bob in the examples given in this section, we also require that the set of bases $\{\mathcal{D}_i\}_{i\in\text{I}_m}$ satisfies the constraint that there exists a maximally entangled state $\lvert \phi\rangle$ of two $l$-dimensional quantum systems $a$ and $a'$ such that $\lvert \phi\rangle$ can be expressed by
\begin{equation}
\label{x5}
\lvert \phi\rangle_{a'a}=\frac{1}{\sqrt{l}}\sum_{r\in\Omega}\bigl\lvert \alpha_r^i\rangle_{a'}\otimes \bigl\lvert \alpha_r^i\rangle_a,
\end{equation}
for all $i\in\text{I}_m$.

From (\ref{x1}) -- (\ref{x5}), we show below that if the protocols of the class defined above do not tolerate any errors in Bob's output then they are $\epsilon_{\text{Bob}}-$secure against dishonest Bob, with
\begin{equation}
\label{5}
\epsilon_{\text{Bob}}= \biggl( \frac{m-1+\sqrt{\lambda}}{m}\biggr)^n.
\end{equation}
Thus, since $\lambda<1$ as given by (\ref{x4}), we have that $\epsilon_{\text{Bob}}\rightarrow 0$ exponentially with $n$, meaning that the protocol is unconditionally secure against dishonest Bob.

We can straightforwardly extend the class of protocols above to tolerate a small fraction of errors. For example, consider protocols with $\Omega=\{0,1\}$, i.e. $\bold{r}_i$ is a string of $n$ bits, for $i\in\text{I}_m$. Bob's output $\bold{r}_c'$ may be considered correct if
$d(\bold{r}_c',\bold{r}_c)\leq n\gamma$ for some small allowed error rate $\gamma\geq 0$, i.e. if the number of bit errors in $\bold{r}_c'$ with respect to $\bold{r}_c$ is not greater than $n\gamma$, for $c\in\text{I}_m$. In this case, we show below that the considered protocols are $\epsilon_{\text{Bob}}^{\gamma}-$secure against dishonest Bob with 
\begin{equation}
\label{5.1}
\epsilon_{\text{Bob}}^{\gamma}=\Biggl[2^{2h(\gamma)}\biggl( \frac{m-1+\sqrt{\lambda}}{m}\biggr)\Biggr]^n,
\end{equation}
for some $\gamma\in(0,\Gamma_{m}^{(\lambda)})$, where $\Gamma_{m}^{(\lambda)}$ is the smallest solution to the following equation
\begin{equation}
\label{5.2}
2^{2h(\Gamma_{m}^{(\lambda)})}\biggl( \frac{m-1+\sqrt{\lambda}}{m}\biggr)=1,
\end{equation}
which satisfies $\Gamma_{m}^{(\lambda)}\leq \frac{1}{2}$, where we recall that $h(\gamma)$ denotes the binary entropy of $\gamma$.
We note from (\ref{5.2}) that since $\gamma<\Gamma_{m}^{(\lambda)}$, the term inside the brackets in (\ref{5.1}) is smaller than unity, hence, $\epsilon_{\text{Bob}}^{\gamma}$ given by (\ref{5.1}) decreases exponentially with $n$. Thus, the protocols are unconditionally secure against dishonest Bob.

We give two specific examples below of protocols for one-out-of-$m$ DQACM of the previous class that satisfy (\ref{x1}) -- (\ref{x5}), from which the security bounds (\ref{5}) and (\ref{5.1}) follow.

\subsection{Example 1}
\label{example1}
We consider the case $m=l=2$ with $\Omega=\{0,1\}$. We set the state $\lvert \phi\rangle=\lvert \Phi^+\rangle$, where $\lvert \Phi^+\rangle$ is the Bell state
\begin{equation}
\label{x6}
\lvert \Phi^+\rangle=\frac{1}{\sqrt{2}}\bigl(\lvert 0\rangle\lvert 0\rangle+\lvert 1\rangle\lvert 1\rangle\bigr).
\end{equation}
Since $l=2$, $\mathcal{D}_0$ and $\mathcal{D}_1$ are qubit orthogonal bases, which without loss of generality we fix on the same plane of the Bloch sphere. Without loss of generality we set $\mathcal{D}_0$ to be the computational basis, given by the states $\lvert \alpha_r^0\rangle=\lvert r\rangle$, for $r\in\{0,1\}$. The basis $\mathcal{D}_1$ is defined by the states
\begin{equation}
\label{x7}
\lvert \alpha_r^1\rangle=(-1)^r\cos\Bigl(\frac{\theta}{2}\Bigr)\lvert r\rangle+\sin\Bigl(\frac{\theta}{2}\Bigr)\lvert \bar{r}\rangle,
\end{equation}
for $r\in\{0,1\}$ and for some $\theta\in\bigl(0,\pi\bigr)$. In this example, we have
\begin{equation}
\label{x7}
\lambda=\max\Bigl\{\cos^2\Bigl(\frac{\theta}{2}\Bigr),\sin^2\Bigl(\frac{\theta}{2}\Bigr)\Bigr\}.
\end{equation}
Since $\theta\in\bigl(0,\pi\bigr)$, we have $\lambda<1$, hence, (\ref{x4}) holds. It is easy to see that $\lvert \alpha_0^1\rangle\lvert \alpha_0^1\rangle+\lvert \alpha_1^1\rangle\lvert \alpha_1^1\rangle=\lvert 0\rangle\lvert 0\rangle+\lvert 1\rangle\lvert 1\rangle$ for any $\theta$, hence, by setting $\lvert \phi\rangle=\lvert\Phi^+\rangle$, (\ref{x5}) holds too. From (\ref{5}), in this example we have that the DQACM protocol is $\epsilon^{\theta}-$secure against dishonest Bob, with
\begin{equation}
\label{xxxxx1}
\epsilon^{\theta}=\biggl( \frac{1+\sqrt{\lambda}}{2}\biggr)^n,
\end{equation}
where $\lambda$ is given by (\ref{x7}). Since $\epsilon^{\theta}$ decreases exponentially with $n$, the DQACM protocol is unconditionally secure against dishonest Bob, for $\theta\in(0,\pi)$.

In this example, in order to enhance the security, it is preferable to have $\theta=\frac{\pi}{2}$, in which case $\mathcal{D}_0$ and $\mathcal{D}_1$ correspond respectively to the computational and Hadamard bases, which are mutually unbiased, giving from (\ref{x7}) the value $\lambda=\frac{1}{2}$. In this case, it follows from (\ref{xxxxx1}) that the DQACM protocol is $\epsilon^{\frac{\pi}{2}}-$secure against dishonest Bob, with
\begin{equation}
\label{xxxxx2}
\epsilon^{\frac{\pi}{2}}=\biggl( \frac{1}{2}+\frac{1}{2\sqrt{2}}\biggr)^n.
\end{equation}

\subsection{Example 2}
\label{example2}

We set arbitrary $m\geq 2$, with $\Omega=\{0,1\}$, hence $l=2$. Since $l=2$, $\mathcal{D}_i$ are qubit orthogonal bases, for $i\in\text{I}_m$. We set the bases to lie on the same plane of the Bloch sphere. Without loss of generality we set this plane to be the $x$-$z$ plane, and we set the basis $\mathcal{D}_0$ to lie on the $z$ axis, i.e. $\mathcal{D}_0$ is the computational basis, which is given by the states $\lvert \alpha_r^0\rangle=\lvert r\rangle$, for $r\in\{0,1\}$. The other bases can be expressed by the states
\begin{equation}
\label{x8}
\lvert \alpha_r^i\rangle=(-1)^r\cos\Bigl(\frac{\theta_i}{2}\Bigr)\lvert r\rangle+\sin\Bigl(\frac{\theta_i}{2}\Bigr)\lvert \bar{r}\rangle,
\end{equation}
for $r\in\{0,1\}$, for different parameters $\theta_i\in\bigl(0,\pi\bigr)$, for $i\in\{1,2,\ldots,m-1\}$, which without loss of generality we order like $\theta_1<\theta_2<\cdots<\theta_{m-1}$.
In this example, we can set
\begin{equation}
\label{x9}
\theta_i=i\frac{\pi}{m},
\end{equation}
for $i\in\{1,2,\ldots,m-1\}$,
which gives
\begin{equation}
\label{x10}
\lambda=\cos^2\Bigl(\frac{\pi}{2m}\Bigr),
\end{equation}
satisfying (\ref{x4}), for $m\geq 2$. As in the Example 1 above, we set $\lvert \phi\rangle=\lvert \Phi^+\rangle$, given by (\ref{x6}), which as above satisfies (\ref{x5}). From (\ref{5}), in this example we have that the DQACM protocol is $\epsilon_m-$secure against dishonest Bob, with
\begin{equation}
\label{xxxxx3}
\epsilon_m=\biggl( \frac{m-1+\cos\bigl(\frac{\pi}{2m}\bigr)}{m}\biggr)^n.
\end{equation}
Since $\epsilon_m$ decreases exponentially with $n$, the DQACM protocol is unconditionally secure against dishonest Bob, for $m\geq 2$.

\subsection{Security against dishonest Bob}

We show below that the class $\mathcal{C}$ of DQACM protocols described in this section satisfying (\ref{x1}) -- (\ref{x5}) are $\epsilon_{\text{Bob}}-$secure against dishonest Bob in the case that errrors in Bob's output are not tolerated, and $\epsilon_{\text{Bob}}^{\gamma}-$secure against dishonest Bob in the case that a small fraction $\gamma$ of errors is tolerated, with $\epsilon_{\text{Bob}}$ and $\epsilon_{\text{Bob}}^{\gamma}$ given by (\ref{5}) and (\ref{5.1}), respectively.

By definition, security against dishonest Bob is analyzed with respect to cheating strategies of the following form. Bob receives the quantum state $\bigl\lvert \Psi_{\bold{r}}^{\bold{s}}\bigr\rangle_A$ from Alice in the quantum system $A$. Bob then applies any quantum operation $O$ on the quantum state $\bigl\lvert \Psi_{\bold{r}}^{\bold{s}}\bigr\rangle_A$, and possibly and ancillary quantum system $E$ or arbitrary finite Hilbert space dimension. Bob then partitions his total system $AE$ into two quantum systems $B_0$ and $B_1$. After receiving $\bold{s}$ from Alice, Bob applies a quantum measurement $\tilde{\text{M}}_0^{\bold{s}}$ on $B_0$ and a quantum measurement $\tilde{\text{M}}_1^{\bold{s}}$ on $B_1$, whose respective outcomes are denoted by $\bold{e}_0$ and $\bold{e}_1$. We show below that the probability $p_n$ that Bob's outputs satisfy $\bold{e}_0=\bold{r}_{l_0}$ and $\bold{e}_1=\bold{r}_{l_1}$ satisfies
\begin{equation}
\label{zzzzz1}
p_n\leq \biggl( \frac{m-1+\sqrt{\lambda}}{m}\biggr)^n,
\end{equation}
for any pair of different numbers $l_0,l_1\in\text{I}_m$. We also show that for Alice's inputs of the form $\bold{r}_0,\bold{r}_1,\ldots,\bold{r}_{m-1}\in\{0,1\}^n$, the probability $p_n'$
that Bob's outputs satisfiy $d(\bold{e}_0,\bold{r}_{l_0})\leq n\gamma$ and $d(\bold{e}_1,\bold{r}_{l_1})\leq n\gamma$ satisfies 
\begin{equation}
\label{zzzzz2}
p_n'\leq 2^{2nh(\gamma)}\biggl( \frac{m-1+\sqrt{\lambda}}{m}\biggr)^n,
\end{equation}
for any pair of different numbers $l_0,l_1\in\text{I}_m$, and for some $\gamma\in(0,\Gamma_{m}^{(\lambda)})$, where $\Gamma_{m}^{(\lambda)}$ is the smallest solution to the  equation
\begin{equation}
\label{zzzzz3}
2^{2h(\Gamma_{m}^{(\lambda)})}\biggl( \frac{m-1+\sqrt{\lambda}}{m}\biggr)=1,
\end{equation}
which satisfies $\Gamma_{m}^{(\lambda)}\leq \frac{1}{2}$. Thus, from (\ref{5}) -- (\ref{5.2}), and from (\ref{zzzzz1}) -- (\ref{zzzzz3}), it follows that the class of DQACM protocols described in this section satisfying (\ref{x1}) -- (\ref{x5}) are $\epsilon_{\text{Bob}}-$secure against dishonest Bob in the case that errors in Bob's output are not tolerated, and $\epsilon_{\text{Bob}}^{\gamma}-$secure against dishonest Bob in the case that a small fraction $\gamma$ of errors is tolerated, with $\epsilon_{\text{Bob}}$ and $\epsilon_{\text{Bob}}^{\gamma}$ given by (\ref{5}) and (\ref{5.1}), respectively.

The most general quantum operation $O$ consists in performing some joint unitary operation $U$, independent of $\bold{s}$ and independent of $\bold{r}$, on the quantum state $\lvert\Psi_\bold{r}^\bold{s}\rangle$ of the quantum system $A$ and a fixed quantum state $\lvert \chi\rangle$ of an ancillary system $E$, which we  assume to be of arbitrary finite Hilbert space dimension, to obtain the state
\begin{equation}
\label{2}
\lvert \Phi_{\bold{r}}^\bold{s}\rangle_{B_0B_1} =U_{AE} \lvert \Psi_{\bold{r}}^\bold{s}\rangle_{A}\lvert \chi\rangle_E ,
\end{equation}
where for simplifying notation we have written $\bold{r}=(\bold{r}_0,\bold{r}_1,\ldots,\bold{r}_{m-1})$, where $AE=B_0B_1$, and where the quantum systems $B_0$ and $B_1$ have arbitrary finite dimensions. Bob partitions the global system $AE$ into two quantum systems $B_0$ and $B_1$. Then, for $i\in\{0,1\}$, Bob applies a projective measurement $\tilde{\text{M}}_{i}^{\bold{s}}=\{\Pi_{i\bold{s}}^{\bold{e}_i}\}_{\bold{e}_i\in\Omega^n}$  on $B_i$ and obtains the outcome $\bold{e}_i$. Bob's cheating probability $p_n$ is given by
\begin{equation}
\label{4}
p_n=\frac{1}{(l)^{mn}(m!)^n}\sum_{\substack{\bold{s}\in\Lambda^{n}\\\bold{r}\in\Omega^{nm}}}\langle \Phi_{\bold{r}}^\bold{s}\rvert \Pi_{0\bold{s}}^{\bold{r}_{l_0}}\otimes \Pi_{1\bold{s}}^{\bold{r}_{l_1}}\rvert \Phi_{\bold{r}}^\bold{s}\rangle,
\end{equation}
where $\bold{r}\in\Omega^{nm}$ denotes that $\bold{r}_i\in\Omega^n$ for $i\in\text{I}_m$, as we have used the notation $\bold{r}=(\bold{r}_0,\bold{r}_1,\ldots,\bold{r}_{m-1})$, and where we recall that $l_0,l_1\in\text{I}_m$ with $l_0\neq l_1$. Then, using (\ref{x1}) -- (\ref{x5}), and in particular, using the property of non-perfect distinguishability of non-orthogonal quantum states exploited in Alice's quantum state preparation (\ref{x1}), as quantified by (\ref{x3}) and (\ref{x4}), we show the bound (\ref{zzzzz1}) below.

From (\ref{zzzzz1}) and (\ref{4}), it is straightforward to derive (\ref{zzzzz2}). Consider a projective measurement $\tilde{\text{M}}_{i}^{\bold{s},\bold{a}}=\{\Pi_{i\bold{s}}^{\bold{e}_i\oplus\bold{a}}\}_{\bold{e}_i\in\Omega^n}$ on $B_i$ for any $n-$bit string $\bold{a}$ in the case $\Omega=\{0,1\}$ and $\lvert\Omega\rvert=l=2$, where we recall that `$\oplus$' denotes bit-wise sum modulo 2. Extending (\ref{4}), Bob's cheating probability $p_n'$ is given by 
\begin{equation}
\label{5.3}
p_n'=\frac{1}{(l)^{mn}(m!)^n}\sum_{\substack{\bold{a}: w(\bold{a}\leq n\gamma)\\ \bold{b}: w(\bold{b}\leq n\gamma)}} \sum_{\substack{\bold{s}\in\Lambda^{n}\\\bold{r}\in\Omega^{nm}}}\langle \Phi_{\bold{r}}^\bold{s}\rvert \Pi_{0\bold{s}}^{\bold{r}_{l_0}\oplus \bold{a}}\otimes \Pi_{1\bold{s}}^{\bold{r}_{l_1}\oplus \bold{b}}\rvert \Phi_{\bold{r}}^\bold{s}\rangle,
\end{equation}
where we recall that $w(\bold{a})$ denotes the Hamming weight of the $n-$bit string $\bold{a}$, i.e. the number of bit entries of $\bold{a}$ equal to `1'. The bound (\ref{zzzzz1}) applies for any pair of projective measurements on $B_0$ and $B_1$, hence, in particular for the projective measurement $\tilde{\text{M}}_{i}^{\bold{s},\bold{a}}=\{\Pi_{i\bold{s}}^{\bold{e}_i\oplus\bold{a}}\}_{\bold{e}_i\in\Omega^n}$ on $B_i$, for $\bold{a}\in\{0,1\}^n$ and $i\in\{0,1\}$. It follows from (\ref{zzzzz1}), (\ref{4}) and (\ref{5.3}) that
\begin{eqnarray}
\label{5.4}
p_n'&\leq& \sum_{\substack{\bold{a}: w(\bold{a}\leq n\gamma)\\ \bold{b}: w(\bold{b}\leq n\gamma)}}\biggl( \frac{m-1+\sqrt{\lambda}}{m}\biggr)^n\nonumber\\
&\leq&2^{2nh(\gamma)}
\biggl( \frac{m-1+\sqrt{\lambda}}{m}\biggr)^n,
\end{eqnarray}
where in the second line we have used that the number of $n-$bit strings $\bold{a}$ with Hamming weight not greater than $n\gamma$ is upper bounded by $2^{nh(\gamma)}$, for $\gamma\leq\frac{1}{2}$, which is shown in section 1.4 of Ref. \cite{Lintbook}, and where $h(\gamma)$ is the binary entropy of $\gamma$. The bound (\ref{zzzzz2}) follows.

\subsubsection{Proof of the bound (\ref{zzzzz1})}

We note that our protocol is mathematically equivalent to the following procedure. First, Alice takes the following actions. She prepares a pair of $l$-dimensional quantum systems $C^j_i$ and $A^j_i$ in the state $\lvert \phi\rangle_{C^j_iA^j_i}$ given by (\ref{x5}) , for $i\in\text{I}_m$ and $j\in[n]$. More precisely, Alice prepares a global quantum system with Hilbert space $C\otimes A$, where $A$ is given by (\ref{x2}), and similarly $C$ is given by 
\begin{equation}
\label{x11}
C=\bigotimes_{\substack{i\in\text{I}_m\\j\in[n]}}C_i^j.
\end{equation}
The quantum system $CA$ is prepared in the quantum state
\begin{equation}
\label{x12}
\lvert \Phi\rangle_{CA}=\bigotimes_{\substack{i\in\text{I}_m\\j\in[n]}}\lvert \phi\rangle_{C_i^jA_i^j}.
\end{equation}
Alice keeps the system $C$ and she sends the system $A$ to Bob. Then, Alice measures $C$ in the orthonormal basis $\bigl\{\bigl\lvert \Psi_{\bold{r}}^{\bold{s}}\bigr\rangle\bigr\}_{\bold{r}\in\Omega^{nm}}$ according to her random value of $\bold{s}\in\Lambda^n$, where $\bold{r}=(\bold{r}_0,\bold{r}_1,\ldots,\bold{r}_{m-1})\in\Omega^{nm}$ means that $\bold{r}_i\in\Omega^n$, for $i\in\text{I}_m$. With probability $l^{-mn}$, Alice measures $\bigl\lvert \Psi_{\bold{r}}^{\bold{s}}\bigr\rangle_C$ and $A$ projects into the state $\lvert \Psi_{\bold{r}}^{\bold{s}}\rangle_{A}$. Bob's unitary operation $U$ in his cheating strategy commutes with Alice's measurements. Thus, we can consider that the global system $CB_0B_1$, before Alice's and Bob's measurement are implemented, is in the state
\begin{equation}
\label{10}
\lvert \Psi\rangle_{CB_0B_1}=\bigl(\mathds{1}_C\otimes U_{B_0B_1}\bigr)\lvert \Phi\rangle_{CA}\lvert \chi\rangle_E,
\end{equation}
where we recall that $B_0B_1=AE$, $E$ is an ancilla, and $B_0$ and $B_1$ have arbitrary finite Hilbert space dimensions. Then, Alice measures $C$ in the orthonormal basis $\bigl\{\bigl\lvert \Psi_{\bold{r}}^{\bold{s}}\bigr\rangle\bigr\}_{\bold{r}\in\Omega^{nm}}$ according to her random value of $\bold{s}\in\Lambda^n$. With probability $l^{-mn}$, Alice measures $\bigl\lvert \Psi_{\bold{r}}^{\bold{s}}\bigr\rangle_C$ and $B_0B_1$ projects into the state $\lvert \Phi_{\bold{r}}^{\bold{s}}\rangle_{B_0B_1}$. After receiving $\bold{s}$, Bob applies the projective measurement $\tilde{\text{M}}_i^\bold{s}$ on $B_i$, for $i\in\{0,1\}$. 

Thus, Bob's cheating probability $p_n$ given by (\ref{4}) equals
\begin{equation}
\label{11}
p_n=\frac{1}{(m!)^n}\sum_{\bold{s}\in\Lambda^n}\text{Tr} \bigl(D_{\bold{s}}\Psi\bigr),
\end{equation}
where $\Psi=\bigl(\lvert \Psi\rangle\langle \Psi\rvert\bigr)_{CB_0B_1}$ and 
\begin{equation}
\label{12}
D_{\bold{s}} = \sum_{\bold{r}\in\Omega^{nm}} \bigl( \bigl\lvert \Psi_{\bold{r}}^{\bold{s}}\bigr\rangle\bigl\langle\Psi_{\bold{r}}^{\bold{s}}\bigr\rvert\bigr)_{C}\otimes \bigl(\Pi_{0\bold{s}}^{\bold{r}_{l_0}}\bigr)_{B_0}\otimes \bigl(\Pi_{1\bold{s}}^{\bold{r}_{l_1}}\bigr)_{B_1},\nonumber
\end{equation}
where we recall that $l_0,l_1\in \text{I}_m$ and $l_0\neq l_1$. 

We derive the bound (\ref{zzzzz1}) with the help of two lemmas of Ref. \cite{TFKW13}. Before stating these lemmas we provide some useful notation. We denote by $\mathcal{H}$ the Hilbert space of the global system $CB_0B_1$, which as said before is arbitrary but finite dimensional. We denote by $\mathcal{L}(\mathcal{H})$ and by $\mathcal{P}(\mathcal{H})$ the sets of linear operators and of positive semi-definite operators on $\mathcal{H}$, respectively. For $A,B\in\mathcal{L}(\mathcal{H})$, the expression $A\geq B$ means that $A-B\in\mathcal{P}(\mathcal{H})$. For $A\in\mathcal{L}(\mathcal{H})$, $\lVert A\rVert$ denotes the Schatten $\infty-$norm of $A$, which gives the largest singular value of $A$, and which coincides with its largest eigenvalue if $A\in\mathcal{P}(\mathcal{H})$.

\begin{lemma}{(Ref. \cite{TFKW13})}
\label{lemma1}
Let $A,B,L\in\mathcal{L}(\mathcal{H})$ such that $A^{\dagger}A\geq B^{\dagger}B$. Then, it holds that $\lVert AL \rVert \geq \lVert BL \rVert$.
\end{lemma}

\begin{lemma}{(Ref. \cite{TFKW13})}
\label{lemma2}
Let $D_1,D_2,\ldots,D_N\in \mathcal{P}(\mathcal{H})$, and let $\{s_k\}_{k\in [N]}$ be a set of $N$ mutually orthogonal permutations of $[N]$. Then
\begin{equation}
\biggl\lVert \sum_{i\in [N]} D_i\biggr \rVert\leq \sum_{k\in [N]}\max_{i\in [N]}\Bigl \lVert \sqrt{D_i}\sqrt{D_{s_k(i)}}\Bigr \rVert.
\end{equation}
\end{lemma}

It follows from Lemma \ref{lemma1} that for $A,A',B,B'\in\mathcal{P}(\mathcal{H})$ satisfying $A'\geq A$ and $B' \geq B$, it holds that $\lVert \sqrt{A'}\sqrt{B'}\rVert \geq \lVert \sqrt{A'}\sqrt{B}\lVert \geq \lVert \sqrt{A}\sqrt{B}\rVert$ \cite{TFKW13}. Thus, if $A,A',B,B'$ are projectors on $\mathcal{H}$ satisfying $A'\geq A$ and $B' \geq B$ then $\lVert A'B'\rVert \geq \lVert AB\rVert$. We use this property below.

To use Lemma \ref{lemma2}, we consider the set of permutations of $\bold{s}\in\Lambda^n$ labeled by $\bold{v}=(v^1,v^2,\ldots,v^n)\in\Lambda^n$ and given by $\bold{s}\rightarrow \bold{s}_{\bold{v}}=(s_{\bold{v}}^1,s_{\bold{v}}^2,\ldots,s_{\bold{v}}^n)$ with $s_{\bold{v}}^j=\bigl(s_{v^j,0}^j,s_{v^j,1}^j,\ldots,s_{v^j,m-1}^j\bigr)$ being a permutation $v^j\in\Lambda$ of $s^j=(s_0^j,s_1^j,\ldots,s_{m-1}^j)$, for $j\in[n]$. This is a set of $(m!)^n$ mutually orthogonal permutations, that is,  $\bold{s}_{\bold{v}}\neq \bold{s}_{\bold{w}}$ if $\bold{v}\neq\bold{w}$, for all $\bold{s}\in\Lambda^n$. To see this, consider a pair of different elements $\bold{v},\bold{w}$ from the set $\Lambda^n$ and any $\bold{s}\in\Lambda^n$. Since $\bold{v}\neq\bold{w}$, there exists at least a $j'\in[n]$ such that $v^{j'}\neq w^{j'}$, hence, $s_{\bold{v}}^{j'}=\bigl(s_{v^{j'},0}^{j'},s_{v^{j'},1}^{j'},\ldots,s_{v^{j'},m-1}^{j'}\bigr)$ and $s_{\bold{w}}^{j'}=\bigl(s_{w^{j'},0}^{j'},s_{w^{j'},1}^{j'},\ldots,s_{w^{j'},m-1}^{j'}\bigr)$ are different permutations of 
$s^{j'}=(s_0^{j'},s_1^{j'},\ldots,s_{m-1}^{j'})$, which means that $s_{\bold{v}}^{j'}\neq s_{\bold{w}}^{j'}$ and therefore that $\bold{s}_{\bold{v}}\neq \bold{s}_{\bold{w}}$.

We have
\begin{eqnarray}
\label{13}
p_n&=&\frac{1}{(m!)^n}\text{Tr}\Biggl(\sum_{\bold{s}\in\Lambda^n}D_{\bold{s}}  \Psi\Biggr)\nonumber\\
&\leq&\frac{1}{(m!)^n} \Biggl\lVert \sum_{\bold{s}\in\Lambda^n} D_{\bold{s}} \Biggr\rVert\nonumber\\
&\leq&\frac{1}{(m!)^n}\sum_{\bold{v}\in\Lambda^n}\max_{\bold{s}\in\Lambda^n}\Bigl\lVert D_{\bold{s}}D_{\bold{s}_{\bold{v}}}\Bigr\rVert,
\end{eqnarray}
where in the first line we used the linearity of the trace, in the second line we used the definition of the Schatten $\infty-$norm, and in the last line we used Lemma \ref{lemma2} and the fact that $D_{\bold{s}}$ and $D_{\bold{s}_{\bold{v}}}$ are projectors. 

In the following we use the notation $\bold{r}=(\bold{r}_0,\bold{r}_1,\ldots,\bold{r}_{m-1})\in\Omega^{nm}$, where $\bold{r}_i\in\Omega^n$, for $i\in\text{I}_{m}$. We define the projectors
\begin{eqnarray}
\label{18}
F_{\bold{s}}&=&\sum_{\bold{r} }\bigl( \bigl\lvert \Psi_{\bold{r}}^{\bold{s}}\bigr\rangle\bigl\langle\Psi_{\bold{r}}^{\bold{s}}\bigr\rvert\bigr)_{C}\otimes \bigl(\Pi_{0\bold{s}}^{\bold{r}_{l_0}}\bigr)_{B_0}\otimes \mathds{1}_{B_1},\nonumber\\
G_{\bold{s}_{\bold{v}}}&=&\sum_{\bold{r}}\bigl( \bigl\lvert \Psi_{\bold{r}}^{\bold{s}_{\bold{v}}}\bigr\rangle\bigl\langle\Psi_{\bold{r}}^{\bold{s}_{\bold{v}}}\bigr\rvert\bigr)_{C}\otimes  \mathds{1}_{B_0}\otimes\bigl(\Pi_{1\bold{s}_{\bold{v}}}^{\bold{r}_{l_1}}\bigr)_{B_1},\nonumber\\
\end{eqnarray}
for $\bold{s},\bold{v}\in\Lambda^n$. We see that $F_{\bold{s}}$ and $G_{\bold{s}_{\bold{v}}}$ satisfy $D_{\bold{s}}\leq F_{\bold{s}}$ and $ D_{\bold{s}_{\bold{v}}}\leq G_{\bold{s}_{\bold{v}}}$, for $\bold{s},\bold{v}\in\Lambda^n$. Thus, we have from Lemma \ref{lemma1} that
\begin{equation}
\label{14}
\lVert D_{\bold{s}}D_{\bold{s}_{\bold{v}}}\rVert^2\leq \lVert F_{\bold{s}}G_{\bold{s}_{\bold{v}}}\rVert ^2=\lVert F_{\bold{s}}G_{\bold{s}_{\bold{v}}}F_{\bold{s}}\rVert,
\end{equation}
where the equality follows from the property $\lVert A\rVert^2=\lVert AA^{\dagger}\rVert=\lVert A^{\dagger}A\rVert$ for any $A\in\mathcal{L}(\mathcal{H})$ \cite{TFKW13} and from the fact that $F_{\bold{s}}$ and $G_{\bold{s}_{\bold{v}}}$ are projectors. Then we show in the Appendix \ref{appendix1} that
\begin{equation}
\label{15}
\lVert F_{\bold{s}}G_{\bold{s}_{\bold{v}}}F_{\bold{s}}\rVert\leq (\lambda)^{\omega_{\bold{v}}},
\end{equation}
where $\omega_{\bold{v}}=\bigl\lvert\{j\in[n]\vert s_{v^j,l_1}^j=s_{l_0}^j\}\bigr\rvert$, that is, $\omega_{\bold{v}}$ is the number of entries $v^j$ of $\bold{v}$ corresponding to a permutation that takes $(a_0,a_1,\ldots,a_{m-1})\in\Lambda$ into $(?,\ldots,?,a_{l_0},?,\ldots,?)\in\Lambda$, where $a_{l_0}$ is in the $l_1$th entry, and where `?' denotes any allowed entry after the permutation. For example, in the case $l_0=0$ and $l_1=1$, $\omega_{\bold{v}}=\bigl\lvert\{j\in[n]\vert s_{v^j,1}^j=s_{0}^j\}\bigr\rvert$ is the number of entries $v^j$ of $\bold{v}$ corresponding to a permutation that takes $(a_0,a_1,\ldots,a_{m-1})\in\Lambda$ into $(?,a_{0},?,\ldots,?)\in\Lambda$, where `?' denotes any allowed entry after the permutation. As explicitly stated by the notation, we see that $\omega_{\bold{v}}$ only depends on $\bold{v}$, but not on $\bold{s}$. Thus, since for a fixed $\bold{v}\in\Lambda^n$, the upper bound on $\lVert F_{\bold{s}}G_{\bold{s}_{\bold{v}}}F_{\bold{s}}\rVert$ given by (\ref{15}) is the same for any $\bold{s}\in\Lambda^n$, we have from (\ref{13}), (\ref{14}) and (\ref{15}) that
\begin{equation}
\label{16}
p_n\leq \frac{1}{(m!)^n}\sum_{\bold{v}\in\Lambda^n} \bigl({\sqrt{\lambda}}\bigr)^{\omega_\bold{v}}.
\end{equation}
We also note that the value of $\omega_{\bold{v}}$ does not depend on the values of $l_0,l_1$, for any pair of different numbers $l_0,l_1\in\text{I}_m$; hence, the bound (\ref{16}) holds for any pair of different numbers $l_0,l_1\in\text{I}_m$.

We compute the sum in (\ref{16}). There are exactly $\bigl(\begin{smallmatrix} n \\ \omega\end{smallmatrix} \bigr)\bigl((m-1)!\bigr)^\omega\bigl(m!-(m-1)!\bigr)^{n-\omega}$ values of $\bold{v}\in\Lambda^n$ satisfying $\omega_\bold{v}=\omega$. We can see this as follows. Consider a $\bold{v}\in\Lambda^n$ such that $\omega_\bold{v}=\omega$. For this $\bold{v}$, there are $\omega$ entries $v^j$ which are permutations of $m$ distinct elements that take $(a_0,a_1,\ldots,a_{m-1})\in\Lambda$ into $(?,\ldots,?,a_{l_0},?,\ldots,?)\in\Lambda$, where $a_{l_0}$ is in the $l_1$th entry. There are $(m-1)!$ possible permutations that take $(a_0,a_1,\ldots,a_{m-1})\in\Lambda$ into $(?,\ldots,?,a_{l_0},?,\ldots,?)\in\Lambda$, where $a_{l_0}$ is in the $l_1$th entry, and $m!-(m-1)!$ that do not. Let us write a $n$-bit string $f_\bold{v}$ whose $j$th entry is $1$ if $v^j$ is a permutation of the form $(a_0,a_1,\ldots,a_{m-1})\rightarrow (?,\ldots,?,a_{l_0},?,\ldots,?)\in\Lambda$, where $a_{l_0}$ is in the $l_1$th entry, or $0$ otherwise. Thus, the number of elements $\bold{v}\in\Lambda^n$ for which $f_\bold{v}$ has $\omega$ entries equal to $1$ and the rest $n-\omega$ entries equal to $0$ is $\bigl(\begin{smallmatrix} n \\ \omega\end{smallmatrix} \bigr)\bigl((m-1)!\bigr)^\omega\bigl(m!-(m-1)!\bigr)^{n-\omega}$. Thus, from (\ref{16}), we have 
\begin{eqnarray}
\label{17}
p_n&\leq&\frac{1}{(m!)^n}\sum_{\omega=0}^{n}\Bigl(\begin{matrix}
  n \\  \omega \end{matrix}\Bigr)\bigl((m-1)!\bigr)^\omega\times\nonumber\\
  &&\qquad\qquad\qquad\times\bigl(m!-(m-1)!\bigr)^{n-\omega}\bigl(\sqrt{\lambda}\bigr)^\omega\nonumber\\
&=&\frac{\bigl((m-1)!\bigr)^n}{(m!)^n}\sum_{\omega=0}^{n}\Bigl(\begin{matrix}
  n \\  \omega \end{matrix}\Bigr)(m-1)^{n-\omega}\bigl(\sqrt{\lambda}\bigr)^\omega\nonumber\\
  &=&\biggl(\frac{m-1}{m}\biggr)^n\sum_{\omega=0}^{n} \Bigl(\begin{matrix}
  n \\  \omega \end{matrix}\Bigr)\biggl(\frac{\sqrt{\lambda}}{m-1}\biggr)^\omega\nonumber\\
  &=&\biggl(\frac{m-1}{m}\biggr)^n\biggl(1+\frac{\sqrt{\lambda}}{m-1}\biggr)^n\nonumber\\
  &=&\biggl(\frac{m-1+\sqrt{\lambda}}{m}\biggr)^n,
  \end{eqnarray}
which is the claimed bound (\ref{zzzzz1}).

\section{A class $\mathcal{P}_{CC}$ of unconditionally secure one-out-of-$m$ SCOT protocols with long-distance classical communication}
\label{section7}

By implementing one-out-of-$m$ DQACM as a subroutine, the following class $\mathcal{P}_{\text{CC}}$ of one-out-of-$m$ SCOT protocols only requires classical  communication among Bob's distant agents. Alice and Bob agree on a reference frame $\mathcal{F}$ in spacetime, on $m$ pairwise spacelike separated output spacetime regions $R_0,R_1,\ldots,R_{m-1}$, and on a spacetime point $Q_i$ of $R_i$, for $i\in\text{I}_m$; they also agree on Alice's message $\bold{x}_i$ in the causal past of $Q_i$ being from the set $\Omega_i=\{0,1\}^{n}$, for some $n\in\mathbb{N}$, and on a maximum  allowed error rate $\gamma_i=\gamma\geq 0$ on Bob's outputs, for $i\in\text{I}_m$. We recall that $G$ is the spacetime region consisting in the intersection of the causal pasts of the spacetime points $Q_0,Q_1,\ldots,Q_{m-1}$. We consider that from Bob's perspective, Alice's inputs $\bold{x}_i\in\{0,1\}^{n}$ are random, for $i\in\{0,1\}^{n}$; and that from Alice's perspective, Bob's input $b\in\text{I}_m=\{0,1,\ldots,m-1\}$ is random.

Alice has trusted agents $\mathcal{A},\mathcal{A}_0,\mathcal{A}_1,\ldots,\mathcal{A}_{m-1}$, and Bob has trusted agents $\mathcal{B},\mathcal{B}_0,\mathcal{B}_1,\ldots,\mathcal{B}_{m-1}$. Each of Alice's (Bob's) agents controls a secure laboratory. It is helpful to consider that $\mathcal{A}$ and $\mathcal{B}$ have adjacent laboratories, and that $\mathcal{A}_i$ and $\mathcal{B}_i$ have adjacent laboratories, for $i\in\text{I}_m$. Alice's (Bob's) agents share secure and authenticated classical channels. There is a classical channel and a quantum channel between $\mathcal{A}$ and $\mathcal{B}$, and there is a classical channel between $\mathcal{A}_i$ and $\mathcal{B}_i$, for $i\in\text{I}_m$. It is possible that $\mathcal{A}$ and $\mathcal{A}_j$ ($\mathcal{B}$ and $\mathcal{B}_j$) are the same agent, for some $j\in\text{I}_m$.

The class $\mathcal{P}_{\text{CC}}$ of one-out-of-$m$ SCOT protocols extends the one-out-of-two  SCOT protocol of Ref. \cite{PGK18.2}. It consists of two stages. Stage I includes quantum communication between the agents $\mathcal{A}$ and $\mathcal{B}$, which can take place within their adjacent laboratories, and which can take an arbitrarily long time, but which must be completed within $G$. For $i\in\text{I}_m$, stage II includes fast classical processing and communication between the agents $\mathcal{A}_i$ and $\mathcal{B}_i$, which can take place within their adjacent laboratories; it also includes classical communication between the -- possibly distant -- pairs of agents $\mathcal{A}$ and $\mathcal{A}_i$, and $\mathcal{B}$ and $\mathcal{B}_i$. The actions performed in the steps 1 to 6 take place within $G$, unless otherwise stated. Fig. \ref{fig4} illustrates the class $\mathcal{P}_{\text{CC}}$ of one-out-of-$m$ SCOT protocols.

\subsection{Stage I}

\begin{enumerate}

\item Alice's agent $\mathcal{A}$ and Bob's agent $\mathcal{B}$ implement the stage I of a one-out-of-$m$ DQACM protocol with random inputs $\bold{s}\in\Lambda_{\text{basis}}$ and $\bold{r}_i\in\{0,1\}^{n}$ by Alice, and a random input $c\in\text{I}_m$ by Bob, for $i\in\text{I}_m$. This consists in $\mathcal{A}$ sending to  $\mathcal{B}$ a quantum state $\lvert\Psi_{\bold{r}}^{\bold{s}}\rangle$ encoding $\bold{r}=(\bold{r}_0,\bold{r}_1,\ldots,\bold{r}_{m-1})$ in a basis labeled by $\bold{s}$, $\mathcal{B}$ applying a quantum measurement $\text{M}_c$ on the received quantum state, and $\mathcal{B}$ obtaining a classical measurement outcome $\bold{d}$. The stage I of the DQACM protocol is completed in the spacetime region $G$. 

\item $\mathcal{A}$ sends copies of $\bold{s}, \bold{r}_0,\bold{r}_1,\ldots,\bold{r}_{m-1}$ to $\mathcal{A}_i$, who receives them in the causal past of $Q_i$, for $i\in\text{I}_m$.

\item $\mathcal{B}$ transmits $c$ and $\bold{d}$ to $\mathcal{B}_i$, who receives these in the causal past of $Q_i$, for $i\in\text{I}_m$.

\end{enumerate}

\subsection{Stage II}
\begin{enumerate}
\setcounter{enumi}{3}
\item Within $G$, $\mathcal{B}$ generates his SCOT input $b\in\text{I}_m$, and transmits the number $b'=b+c$ modulo $m$ to $\mathcal{A}$, who receives it within $G$.

\item For $i\in\text{I}_m$, $\mathcal{B}$ transmits $b$ to $\mathcal{B}_i$, who receives it in the causal past of $Q_i$.

\item For $i\in\text{I}_m$, $\mathcal{A}$ transmits $b'$ to $\mathcal{A}_i$, who receives it in the causal past of $Q_i$.

\item For $i\in\text{I}_m$, $\mathcal{A}_i$ generates $\bold{x}_i$ in the causal past of $Q_i$, and gives $\bold{t}_i=\bold{r}_{b'-i}\oplus \bold{x}_{i}$ to $\mathcal{B}_i$ at $Q_i$, where $b'-i$ is modulo $m$.

\item For $i\in\text{I}_m$, $\mathcal{A}_i$ gives $\bold{s}$ to $\mathcal{B}_i$ at $Q_i$. This corresponds to the first step in stage II of the DQACM protocol.

\item Within the spacetime region $R_b$, $\mathcal{B}_b$ uses $\bold{s}$, $\bold{d}$ and $c$ to obtain the output $\bold{r}_c$ (or $\bold{r}_c'$ close to $\bold{r}_c$ according to a predetermined threshold) of the DQACM protocol. This corresponds to the second step in stage II of the DQACM protocol.

\item Within $R_b$, $\mathcal{B}_b$ outputs $\bold{x}_b=\bold{r}_c\oplus \bold{t}_b$ (or $\bold{x}_b'=\bold{r}_c'\oplus \bold{t}_b$, which is close to $\bold{x}_b$ according to a predetermined threshold).

\end{enumerate}

\begin{figure}
\includegraphics{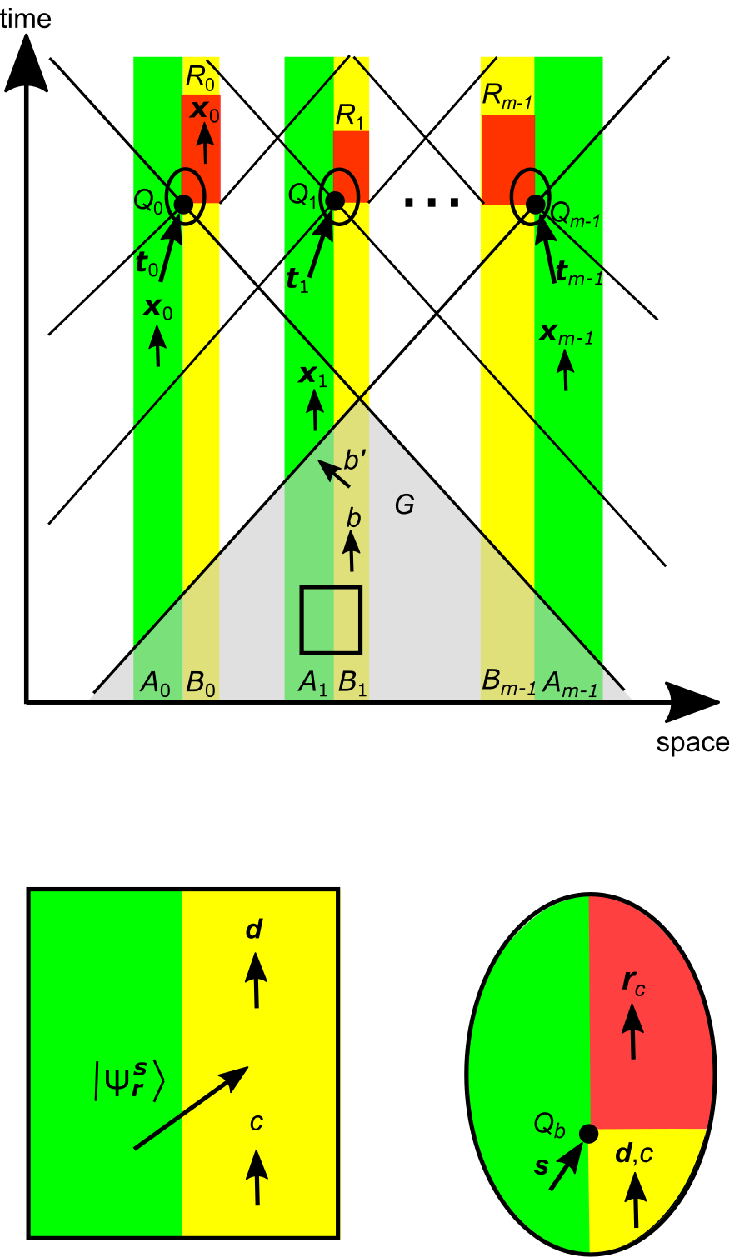}
 \caption{\label{fig4} Illustration of the class $\mathcal{P}_{\text{CC}}$ of one-out-of-$m$ SCOT protocols in a two-dimensional spacetime diagram in a frame $\mathcal{F}$ of Minkowski spacetime. The world lines of the laboratories of Alice's agents $\mathcal{A}_0,\mathcal{A}_1,\ldots,\mathcal{A}_{m-1}$ (green rectangles), and of the laboratories of Bob's agents $\mathcal{B}_0,\mathcal{B}_1,\ldots,\mathcal{B}_{m-1}$ (yellow rectangles) are indicated. The small dots represent the spacetime points $Q_0,Q_1,\ldots,Q_{m-1}$. The thin diagonal lines represent light rays. The spacetime region $G$, consisting in the intersection of the causal pasts of $Q_0,Q_1,\ldots,Q_{m-1}$, is represented by the grey shaded area. The spacetime regions $R_i$, where Bob's agents must obtain Alice's inputs $\bold{x}_i$, correspond to the small red rectangles, for $i\in\{0,1,\ldots,m-1\}$. $R_0,R_1,\ldots,R_{m-1}$ are pairwise spacelike separated. Alice's (Bob's) inputs and outputs obtained within her (his) laboratories are represented by vertical arrows. Communication from Alice to Bob, or vice versa, is represented by diagonal arrows. Top: Stage I of the one-out-of-$m$ DQACM protocol is completed within $G$ (black-edged square); and stage II takes place around the spacetime points $Q_0,Q_1,\ldots,Q_{m-1}$ (black-edged ellipses). In the illustrated example, we take Bob's (Alice's) agents $\mathcal{B}$ ($\mathcal{A}$) and $\mathcal{B}_1$ ($\mathcal{A}_1$) to be the same agent, and we take $b=0$. Lower left: Stage I of the one-out-of-$m$ DQACM protocol. Lower right: Stage II of the one-out-of-$m$ DQACM protocol, which can be completed around any plurality of the spacetime points $Q_0,Q_1,\ldots,Q_{m-1}$ that include the spacetime point $Q_b$.}
 \end{figure}

\subsection{Comments and variations}

We note that the class $\mathcal{P}_{\text{CC}}$ of one-out-of-$m$ SCOT protocols
shares various important properties with the one-out-of-two SCOT protocol of Ref. \cite{PGK18.2}. First, $\mathcal{B}$ has the freedom to choose $b$ after he has measured the quantum state received from $\mathcal{A}$. Thus, the quantum communication and quantum measurement steps can take an arbitrarily long time, but they must be completed within $G$. Second, Alice has the freedom to choose her inputs $\bold{x}_0$, $\bold{x}_1,\ldots,\bold{x}_{m-1}$ in real time, i.e. $\mathcal{A}_i$ can generate $\bold{x}_i$ anywhere in the causal past of $Q_i$, for $i\in\text{I}_m$. Third, different variations of the protocols can be considered. For example, if $\mathcal{B}$ does not send $b$ to $\mathcal{B}_i$, $\mathcal{B}_i$ can act assuming that $b=i$, for $i\in\text{I}_m$. In particular, for $i\in\text{I}_m$, independently of whether $\mathcal{B}$ sends $b$ to $\mathcal{B}_i$, $\mathcal{B}_i$ can output a message $\bold{r}_c'$ that is equal to (or close to) Alice's input $\bold{r}_c$ (see Fig. \ref{fig4}); although this does not allow $\mathcal{B}_i$ to obtain $\bold{x}_i$ (or a message $\bold{x}_i'$ close to $\bold{x}_i$, unless $i=b$) as shown below.

\subsection{Correctness}
We assume that the DQACM subroutine is $\epsilon_{\text{cor}}-$correct in the case that 
an error rate $\gamma\geq 0$ on Bob's output $\bold{r}_c'$ is tolerated, for $c\in\text{I}_m$. This means that $d(\bold{r}_c',\bold{r}_c)\leq \gamma n$ with probability not smaller than $1-\epsilon_{\text{cor}}$, for $c\in\text{I}_m$. Therefore, $d(\bold{x}_b',\bold{x}_b) \leq \gamma n$ with probability not smaller than $1-\epsilon_{\text{cor}}$, for $b\in\text{I}_m$. It follows that the one-out-of-$m$ SCOT protocols from the class $\mathcal{P}_{\text{CC}}$ are $\epsilon_{\text{cor}}-$correct in the case that an error rate $\gamma\geq 0$ on Bob's output $\bold{x}_b'$ is tolerated, for $b\in\text{I}_m$.

\subsection{Security against dishonest Alice}

Neither in the DQACM subroutine, nor in the whole one-out-of-$m$ SCOT protocol, Bob gives Alice any physical systems. Thus, Alice cannot obtain any information about Bob's SCOT input $b\in\text{I}_m$. It follows that the SCOT protocol is perfectly secure against dishonest Alice.

\subsection{Security against dishonest Bob}

We assume that the DQACM subroutine is $\epsilon_{\text{Bob}}-$secure against dishonest Bob. We show that the one-out-of-$m$ SCOT protocol from the class $\mathcal{P}_{\text{CC}}$ is $\epsilon_{\text{Bob}}-$secure against dishonest Bob. It follows that if the DQACM subroutine is unconditionally secure against dishonest Bob, i.e. if $\epsilon_{\text{Bob}}$ goes to zero by increasing the number of bits $n$ of Alice's input messages and possibly some other security parameters, then the SCOT protocol is unconditionally secure.

In order to show security against dishonest Bob, we assume that Alice follows the 
one-out-of-$m$ SCOT protocol honestly and Bob applies an arbitrary cheating strategy allowed by quantum theory and relativity. Consider a general cheating strategy by Bob in which he outputs a message $\bold{x}_i'\in\{0,1\}^n$ in $R_i$ and a message $\bold{x}_j'\in\{0,1\}^n$ in $R_j$, which in a successful cheating strategy are equal to -- or very close to -- Alice's inputs $\bold{x}_i$ and $\bold{x}_j$, respectively, for some pair of different numbers $i,j\in\text{I}_m$. Given that Alice's inputs $\bold{r}_0,\bold{r}_1,\ldots,\bold{r}_{m-1}\in\{0,1\}^n$ are random, and that $\mathcal{A}_i$ gives the message $\bold{t}_i=\bold{r}_{ b'-i}\oplus\bold{x}_i$ to $\mathcal{B}_i$ at $Q_i$, and $\mathcal{A}_j$ gives the message $\bold{t}_j=\bold{r}_{b'-j}\oplus\bold{x}_j$ to $\mathcal{B}_j$ at $Q_j$, the goal of Bob's cheating strategy is that $\mathcal{B}_i$ and $\mathcal{B}_j$ obtain respective strings $\bold{r}_{b'-i}'$ and $\bold{r}_{b'-j}'$, in $R_i$ and $R_j$, that are equal to -- or close to -- $\bold{r}_{b'-i}$ and $\bold{r}_{b'-j}$ with high probability, so that $\mathcal{B}_i$ outputs $\bold{x}_i'=\bold{r}_{b'-i}'\oplus\bold{t}_i$ in $R_i$ and $\mathcal{B}_j$ outputs $\bold{x}_j'=\bold{r}_{b'-j}'\oplus\bold{t}_j$ in $R_j$, which are equal to -- or close to -- $\bold{x}_i$ and $\bold{x}_j$, respectively, with high probability.

Therefore, Bob's general strategy consists of three main steps. In the first step, Bob's agent $\mathcal{B}$ receives the quantum state $\lvert\Psi_\bold{r}^\bold{s}\rangle$ in a quantum system $A$ from Alice's agent $\mathcal{A}$ and applies a quantum operation $O$ on $A$ and an extra ancillary system $E$ consisting in a unitary operation $U$ on $AE$, independent of $\bold{r}$ and independent of $\bold{s}$, producing two quantum systems $B_0$ and $B_1$, including also a measurement $\tilde{\text{M}}'$ producing a classical outcome $(b',i,j)\in\Gamma$ encoded in a system $B''$, where we define $\Gamma=\{(k,i,j)\in \text{I}_m\times \text{I}_m\times \text{I}_m\vert i\neq j\}$. $\mathcal{B}$ then sends $b'$ to $\mathcal{A}$ (who sends $b'$ to $\mathcal{A}_k$, for $k\in\text{I}_m$) and he sends the classical message $(b',i,j)$ encoded in a classical system $B_0''$ and the quantum system $B_0$ to Bob's agent $\mathcal{B}_i$, and the classical message $(b',i,j)$ encoded in a classical system $B_1''$ and the quantum system $B_1$ to Bob's agent $\mathcal{B}_j$.
Thus, $\mathcal{B}$ sends $B_0B_0''$ to $\mathcal{B}_i$ and $B_1B_1''$ to $\mathcal{B}_j$, while $B''$ is held by $\mathcal{B}$, except for $b'$, which $\mathcal{B}$ sends to $\mathcal{A}$.

In the second step, after reception of $\bold{s}$ from $\mathcal{A}_i$, and after reception of the classical message $(b',i,j)$ and of the quantum system $B_0$ from $\mathcal{B}$, $\mathcal{B}_i$ applies a quantum measurement $\tilde{\text{M}}_{0,b',i,j}^\bold{s}$ -- depending on both $\bold{s}$ and $(b',i,j)$ -- on $B_0$ and obtains the guess $\bold{r}_{b'-i}'$ of $\bold{r}_{b'-i}$. Similarly, after reception of $\bold{s}$ from $\mathcal{A}_j$, and after reception of the classical message $(b',i,j)$ and of the quantum system $B_1$ from $\mathcal{B}$, $\mathcal{B}_j$ applies a quantum measurement $\tilde{\text{M}}_{1,b',i,j}^\bold{s}$  on $B_1$ and obtains the guess $\bold{r}_{b'-j}'$ of $\bold{r}_{b'-j}$.

As shown in the Appendix \ref{appendix2}, the two steps above are mathematically equivalent to the following situation. More precisely, for any pair of different numbers $i,j\in\text{I}_m$, we show that the joint probability that Bob's agent $\mathcal{B}_i$ obtains a particular outcome $\bold{r}_{b'-i}'$ as his guess of $\bold{r}_{b'-i}$ and Bob's agent $\mathcal{B}_j$ obtains a particular outcome $\bold{r}_{b'-j}'$ as his guess of $\bold{r}_{b'-j}$ in the procedure of the two steps above is the same in the procedure described in the paragraph below.

Bob's agent $\mathcal{B}$ applies a quantum operation $O'$ on the received quantum state $\lvert\Psi_\bold{r}^\bold{s}\rangle_A$ and an extra ancillary system $E'=EB_0''B_1''B''$, producing two quantum systems $B_0'=B_0B_0''B''$ and $B_1'=B_1B_1''$. The operation $O'$ consists in $\mathcal{B}$ applying the unitary operation $U$ on $AE$ of the quantum operation $O$ above, partitioning $AE$ into two subsystems $B_0$ and $B_1$, applying the quantum measurement $\tilde{\text{M}}'$ on $B_0B_1$ and preparing each of the quantum systems $B_0''$, $B_1''$ and $B''$ in a quantum state $\lvert \mu_{b',i,j}\rangle$, conditioned on the outcome of $\tilde{\text{M}}'$ being $(b',i,j)$, for $(b',i,j)\in\Gamma$, where $\{\lvert \mu_{b',i,j}\rangle\}_{(b',i,j)\in\Gamma}$ is an orthonormal basis of each of the quantum systems $B_0''$, $B_1''$ and $B''$. Conditioned on the outcome of $\tilde{\text{M}}'$ being $(b',i,j)$, $\mathcal{B}$ sends $b'$ to Alice's agent $\mathcal{A}$ in part of the system $B''$, and  $\mathcal{B}$ sends the joint system $B_0B_0''$ ($B_1B_1''$) to Bob's agent $\mathcal{B}_i$ ($\mathcal{B}_j$). A quantum measurement $\tilde{\text{M}}_{0}^{\bold{s}}$ is applied on the joint system $B_0'=B_0B_0''B''$, with $\mathcal{B}_i$ obtaining the outcome $r_{b'-i}'$ from $B_0B_0''$, which is his guess of $\bold{r}_{b'-i}$. Bob's agent $\mathcal{B}_j$ applies a quantum measurement $\tilde{\text{M}}_{1}^{\bold{s}}$ on the joint system $B_1'=B_1B_1''$ and obtains a classical outcome $r_{b'-j}'$, which is his guess of $\bold{r}_{b'-j}$.


Finally, in the third step, after reception of the message $\bold{t}_i$ from $\mathcal{A}_i$, $\mathcal{B}_i$ computes his guess $\bold{x}_i'=\bold{r}_{b'-i}'\oplus\bold{t}_i$ of $\bold{x}_i$ and outputs it in $R_i$. Similarly, after reception of $\bold{t}_j$ from $\mathcal{A}_j$, $\mathcal{B}_j$ computes his guess $\bold{x}_j'=\bold{r}_{b'-j}'\oplus\bold{t}_j$ of $\bold{x}_j$ and outputs it in $R_j$. 

By assumption, the DQACM subroutine is $\epsilon_{\text{Bob}}-$secure against dishonest Bob. By definition, since we assume that Alice follows the protocol honestly, for any pair of different numbers $k$ and $l$ from the set $\text{I}_m$, for any quantum operation $O'$ independent of $\bold{r}$ and independent of $\bold{s}$ applied by Bob on the quantum state $\lvert \Psi_ {\bold{r}}^{\bold{s}}\rangle_A$ that produces at least two quantum systems $B_0'$ and $B_1'$, and for any sets of quantum measurements $\bigl\{\tilde{\text{M}}_0^\bold{s}\bigr\}_{\bold{s}\in\Lambda_{\text{basis}}}$ and $\bigl\{\tilde{\text{M}}_1^\bold{s}\bigr\}_{\bold{s}\in\Lambda_{\text{basis}}}$, the probability to obtain respective outcomes $\bold{r}_k'$ and $\bold{r}_l'$ satisfying $d(\bold{r}_k',\bold{r}_k)\leq \gamma n$ and $d(\bold{r}_l',\bold{r}_l)\leq \gamma n$, by respectively applying $\tilde{\text{M}}_0^\bold{s}$ on $B_0'$ and $\tilde{\text{M}}_1^\bold{s}$ on $B_1'$, is not greater than $\epsilon_{\text{Bob}}$. Thus, by considering $k=b'-i$ and $l=b'-j$, we see from the first and second steps of Bob's general cheating strategy in the one-out-of-$m$ SCOT protocol, that the probability that Bob's agents $\mathcal{B}_i$ and $\mathcal{B}_j$ output $\bold{r}_{b'-i}'$ in $R_i$ and $\bold{r}_{b'-j}'$ in $R_j$ satisfying $d(\bold{r}_{b'-i}',\bold{r}_{b'-i})\leq \gamma n$ and $d(\bold{r}_{b'-j}',\bold{r}_{b'-j})\leq \gamma n$, respectively, is not greater than $\epsilon_{\text{Bob}}$. Since in the third step of Bob's cheating strategy $\mathcal{B}_i$ outputs $\bold{x}_i'=\bold{r}_{b'-i}'\oplus\bold{t}_i$ in $R_i$ and $\mathcal{B}_j$ outputs $\bold{x}_j'=\bold{r}_{b'-j}'\oplus\bold{t}_j$ in $R_j$, and since $\bold{x}_i=\bold{r}_{b'-i}\oplus\bold{t}_i$ and $\bold{x}_j=\bold{r}_{b'-j}\oplus\bold{t}_j$, it follows that the probability that 
Bob's agents $\mathcal{B}_i$ and $\mathcal{B}_j$ output $\bold{x}_{i}'$ in $R_i$ and $\bold{x}_{j}'$ in $R_j$ satisfying $d(\bold{x}_{i}',\bold{x}_{i})\leq \gamma n$ and $d(\bold{x}_{j}',\bold{x}_{j})\leq \gamma n$, respectively, is not greater than $\epsilon_{\text{Bob}}$. This means, by definition, that the one-out-of-$m$ SCOT protocol is $\epsilon_{\text{Bob}}-$secure against dishonest Bob, as claimed.

\subsection{Examples}

We consider that the DQACM subroutine belongs to the class $\mathcal{C}$ introduced in section \ref{section6} with $l=2$, i.e. with inputs by Alice $\bold{r}_i\in\{0,1\}^n$ for $i\in\text{I}_m$. We consider separately the case where no errors in Bob's output are tolerated and the case where a small fraction of errors is tolerated in Bob's outputs.

We consider first the ideal case of no errors. In this case, the DQACM subroutine is perfectly correct, i.e $0-$correct. It follows that a one-out-of-$m$ SCOT protocol of the class $\mathcal{P}_{\text{CC}}$ using this DQACM subroutine is perfectly correct in the ideal case of no errors.

The class $\mathcal{C}$ of DQACM protocols with $l=2$ is $\epsilon_{\text{Bob}}-$secure against dishonest Bob, with $\epsilon_{\text{Bob}}$ given by (\ref{5}), hence, unconditionally secure against dishonest Bob, as $\epsilon_{\text{Bob}}$ decreases exponentially with $n$. It follows that a one-out-of-$m$ SCOT protocol of the class $\mathcal{P}_{\text{CC}}$ using this DQACM subroutine is $\epsilon_{\text{Bob}}-$secure against dishonest Bob, with $\epsilon_{\text{Bob}}$ given by (\ref{5}), hence, unconditionally secure against dishonest Bob. For example, consider that the DQACM subroutine is given by the protocol of Example 2 in section \ref{example2}. In this case, the DQACM protocol is $\epsilon_m-$secure against dishonest Bob, with $\epsilon_m$ given by (\ref{xxxxx3}). Thus, from the arguments above, the one-out-of-$m$ SCOT protocol is $\epsilon_m-$secure against dishonest Bob, with $\epsilon_m$ given by (\ref{xxxxx3}). Since $\epsilon_m$ decreases exponentially with $n$, the SCOT protocol is unconditionally secure against dishonest Bob.

Now we consider the case where a small fraction of errors is tolerated in Bob's outputs. In the case that the fraction of bit errors in Bob's output in the DQACM subroutine is below a threshold $\gamma''\geq 0$, the DQACM subroutine is perfectly correct by setting the allowed error rate $\gamma'$ equal or greater than $\gamma''$. Thus, the SCOT protocol of the class $\mathcal{P}_{\text{CC}}$ using this DQACM subroutine is perfectly correct in the case that a maximum allowed error rate $\gamma$ on Bob's SCOT outputs is set to a value equal or greater than $\gamma'$.

The class $\mathcal{C}$ of DQACM protocols with $l=2$ is $\epsilon_{\text{Bob}}^\gamma-$secure against dishonest Bob, with $\epsilon_{\text{Bob}}^\gamma$ given by (\ref{5.1}), hence, unconditionally secure against dishonest Bob, as $\epsilon_{\text{Bob}}^\gamma$ decreases exponentially with $n$, if $\gamma \leq \Gamma_m^{(\lambda)}$, where $\Gamma_m^{(\lambda)}$ is the smallest solution of the equation (\ref{5.2}). It follows that a one-out-of-$m$ SCOT protocol of the class $\mathcal{P}_{\text{CC}}$ using this DQACM subroutine is $\epsilon_{\text{Bob}}^\gamma-$secure against dishonest Bob, with $\epsilon_{\text{Bob}}^{\gamma}$ given by (\ref{5.1}), hence, unconditionally secure against dishonest Bob, if $\gamma \leq \Gamma_m^{(\lambda)}$, where $\Gamma_m^{(\lambda)}$ is the smallest solution of the equation (\ref{5.2}).

\section{Generalizations}
\label{section8}

We note that the one-out-of-$m$ SCOT protocols of the class $\mathcal{P}_{\text{CC}}$ use one-out-of-$m$ DQACM protocols as a fundamental primtive. In the one-out-of-$m$ DQACM protocols, Alice's agent $\mathcal{A}$ encodes $m$ random messages $\bold{r}_0,\bold{r}_1,\ldots,\bold{r}_{m-1}$ from the agreed sets in a quantum state that she gives to Bob's agent $\mathcal{B}$ in a spacetime region $G$, in such a way that the probability that Bob obtains Alice's input $\bold{r}_i$ -- or a message $\bold{r}_i'$ very close to $\bold{r}_i$ -- in a spacetime region $R$, and also Alice's input $\bold{r}_j$ -- or a message $\bold{r}_j'$ very close to $\bold{r}_j$ -- in a spacetime region $R'$ that is spacelike separated from $R$, is very small, for any pair of different numbers $i,j$ from the set $\text{I}_m$. We can then use the one-out-of-$m$ DQACM primitive to consider more general  SCOT schemes, as we illustrate below. We can also extend the definition of DQACM, which allows us to further generalize the definition of SCOT.

\subsection{Using a one-out-of-$m$ DQACM subroutine to implement generalized versions of SCOT}

Consider a more general setting for SCOT in a spacetime that is Minkowski or close to Minkowski. Alice and Bob agree on a reference frame $\mathcal{F}$ in spacetime. Alice and Bob agree on $M\geq 2$ pairwise spacelike separated output spacetime regions $R_0,R_1,\ldots,R_{M-1}$, and on a spacetime point $Q_i$ of $R_i$, for $i\in\text{I}_M$. Alice inputs messages $\bold{x}_i^j\in\Omega_i^j$ in the causal past of $Q_i$, for $j\in[N_i]$, where the set $\Omega_i^j$ and the number $N_i\in\mathbb{N}$ are previously agreed by Alice and Bob, for $i\in\text{I}_M$.

For some previously agreed integer $m\geq2$, Alice and Bob perform a one-out-of-$m$ DQACM subroutine. Alice encodes random messages $\bold{r}_0,\bold{r}_1,\ldots,\bold{r}_{m-1}$ from sets previously agreed with Bob in a quantum state $\lvert \Psi_\bold{r}^\bold{s}\rangle$, where $\bold{s}$ denotes a basis randomly chosen by Alice from a set $\Lambda_{\text{basis}}$ of non-mutually orthogonal bases previously agreed with Bob, and where we denote $\bold{r}=(\bold{r}_0,\bold{r}_1,\ldots,\bold{r}_{m-1})$. Bob receives the quantum state  $\lvert \Psi_\bold{r}^\bold{s}\rangle$ from Alice in the spacetime region $G$, which is the intersection of the causal pasts of $Q_0,Q_1,\ldots,Q_{M-1}$. Bob inputs a number $b\in\text{I}_m$ in $G$ and obtains a message $\bold{r}_b$ of his choice in any number of the output spacetime regions. Alice applies a classical encoding $\bold{t}_i^j=\mathcal{E} (\bold{x}_i^j,\bold{r}_{l_i^j})$ of $\bold{x}_i^j$ using $\bold{r}_{l_i^j}$, with the encoding being previously agreed with Bob, for some $l_i^j\in\text{I}_m$, for $j\in[N_i]$ and $i\in\text{I}_M$. For example, if $\bold{x}_i^j,\bold{r}_{l_i^j}\in\{0,1\}^{n_i^j}$ for some $n_i^j\in\mathbb{N}$, then we can set $\bold{t}_i^j=\bold{x}_i^j\oplus \bold{r}_{l_i^j}$. Alice then gives $\bold{s}$ to Bob in $Q_i$, for $i\in\text{I}_M$. Bob is then able to complete the DQACM protocol and obtain the message $\bold{r}_b$ of his choice (or a message $\bold{r}_b'$ close to $\bold{r}_b$) in any number of the output spacetime regions. Alice also gives the encoding message $\bold{t}_i^j$ to Bob at $Q_i$, for $j\in[N_i]$ and $i\in\text{I}_M$. Bob is then able to decode Alice's message $\bold{x}_i^j=\mathcal{D}(\bold{t}_i^j,\bold{r}_{l_i^j})$ (or a message ${\bold{x}_i^j}'=\mathcal{D}(\bold{t}_i^j,\bold{r}_{l_i^j}')$ close to $\bold{x}_i^j$) in the output spacetime region $R_i$ via a decoding  $\mathcal{D}$ using $\bold{t}_i^j$ and $\bold{r}_b$ (or $\bold{r}_b'$) if $b=l_i^j$. For example, if $\bold{x}_i^j,\bold{r}_{l_i^j}\in\{0,1\}^{n_i^j}$ for some $n_i^j\in\mathbb{N}$ and $\bold{t}_i^j=\bold{x}_i^j\oplus \bold{r}_{l_i^j}$ then Bob computes $\bold{x}_i^j=\bold{t}_i^j\oplus \bold{r}_{l_i^j}$ (or ${\bold{x}_i^j}'=\bold{t}_i^j\oplus \bold{r}_{l_i^j}'$).

The security guarantee of the one-out-of-$m$ DQACM subroutine is that Bob cannot obtain with non-negligible probability Alice's input $\bold{r}_i$ -- or a message $\bold{r}_i'$ very close to $\bold{r}_i$ -- in one output spacetime region and also Alice's input $\bold{r}_j$ -- or a message $\bold{r}_j'$ very close to $\bold{r}_j$ -- in another output spacetime region, for any pair of different numbers $i,j$ from the set $\text{I}_m$. Thus, with unconditional security, it is guaranteed in this generalized version of SCOT that Bob cannot obtain a message $\bold{x}_i^j$ -- or a message ${\bold{x}_i^j}'$ very close to $\bold{x}_i^j$ -- in the output spacetime region $R_i$ and a message $\bold{x}_k^h$ -- or a message ${\bold{x}_k^h}'$ very close to $\bold{x}_k^h$ -- in the output spacetime region $R_k$, for any pair of different numbers $i,k$ from the set $\text{I}_M$ for which it holds that $l_i^j\neq l_k^h$. Therefore, in order to satisfy specific security constraints, Alice and Bob must previously agree on the classical encodings $\bold{t}_i^j=\mathcal{E}(\bold{x}_i^j,\bold{r}_{l_i^j})$ and decodings $\mathcal{D}(\bold{t}_i^j,\bold{r}_{l_i^j})$, and particularly on the messages $\bold{r}_{l_i^j}$ of these encodings in order to satisfy the desired security conditions.

\subsection{$k$-out-of-$m$ DQACM and SCOT}

We can consider generalizations of one-out-of-$m$ DQACM to a $k$-out-of-$m$ setting for arbitrary natural numbers $k<m$ and $m\geq 2$. Broadly speaking, a $k$-out-of-$m$ DQACM protocol involves the following steps. Alice prepares a quantum state $\lvert \Psi_{\bold{r}}^{\bold{s}}\rangle$ that she gives to Bob, where $\bold{r}=(\bold{r}_0,\bold{r}_1,\ldots,\bold{r}_{m-1})$ and $\bold{s}$ are randomly chosen by Alice from predetermined sets $\Omega_{\text{outcome}}$ and $\Lambda_{\text{basis}}$, respectively. Bob chooses $k$ different numbers $c_0,c_1,\ldots,c_{k-1}$ from the set $\text{I}_m$ and applies a quantum measurement $\text{M}_c$ labeled by $c=(c_0,c_1,\ldots,c_{k-1})$ on the quantum state $\lvert \Psi_{\bold{r}}^{\bold{s}}\rangle$ and obtains a classical measurement outcome $\bold{d}$. Alice then gives $\bold{s}$ to Bob. Bob then uses $c$, $\bold{s}$ and $\bold{d}$ to decode Alice's inputs $\bold{r}_{c_0},\bold{r}_{c_1},\ldots,\bold{r}_{c_{k-1}}$. A $k$-out-of-$m$ DQACM protocol must satisfy a security condition against dishonest Bob, according to which, for any subset $\{l_i\}_{i=0}^k$ of $k+1$ different elements from the set $\text{I}_m$, for any quantum operation $O$ independent of $\bold{r}$ and independent of $\bold{s}$ applied by Bob on the received quantum state $\lvert \Psi_{\bold{r}}^{\bold{s}}\rangle$ that produces at least $k+1$ quantum systems $B_0,B_1,\ldots,B_k$, and for any quantum measurement $\tilde{\text{M}}_{i}^{\bold{s}}$ applied on $B_i$ depending on $\bold{s}$, the probability that the measurement outcome $\bold{r}_{l_i}'$ is equal to $\bold{r}_{l_i}$ - or very close to $\bold{r}_{l_i}$ according to a predetermined threshold -- for all $i\in\{0,1,\ldots,k\}$ is not greater than a small security bound $\epsilon>0$, which ideally decreases by increasing the size of Alice's input messages and possibly by increasing some other security parameters. 

We can then use a $k$-out-of-$m$ DQACM protocol as a fundamental primitive to implement
more general SCOT protocols. For example, we may consider the following definition of $k$-out-of-$m$ SCOT, for natural numbers $k<m$ and $m\geq 2$. We may consider that spacetime is Minkowski or close to Minkowski. Alice and Bob agree on a reference frame $\mathcal{F}$ in spacetime. Alice and Bob agree on $m\geq 2$ pairwise spacelike separated output spacetime regions $R_0,R_1,\ldots,R_{m-1}$, and on a spacetime point $Q_i$ of $R_i$, for $i\in\text{I}_m$. Alice inputs a message $\bold{x}_i\in\Omega_i$ in the causal past of $Q_i$, where the set $\Omega_i$ is previously agreed by Alice and Bob, for $i\in\text{I}_m$. Bob inputs $k$ different numbers $b_0,b_1,\ldots,b_{k-1}$ from the set $\text{I}_m$. In a correct $k$-out-of-$m$ SCOT protocol Bob outputs $\bold{x}_{b_i}$ -- or a message very close to it according to a predetermined threshold-- in $R_{b_i}$, for $i\in\text{I}_k$. We may define security against dishonest Alice as the guarantee that Alice cannot obtain any information about Bob's input $b=(b_0,b_1,\ldots,b_{k-1})$ anywhere in spacetime, when Bob follows the honest protocol and Alice implements an arbitrary cheating strategy allowed by quantum theory and relativity. We define a $k$-out-of-$m$ SCOT protocol to be secure against dishonest Bob if, when Alice follows the protocol honestly and Bob implements an arbitrary cheating strategy allowed by quantum theory and relativity, the probability that Bob outputs $\bold{x}_i$ -- or a message very close to $\bold{x}_i$ according to a predetermined threshold -- in $R_i$, for any $k+1$ different numbers $i$ from the set $\text{I}_m$, is not greater than a small security bound $\epsilon>0$, which ideally decreases by increasing the size of Alice's input messages and possibly by increasing some other security parameters. 

We outline a way to construct $k$-out-of-$m$ DQACM protocols by extending the one-out-of-$m$ DQACM protocols of the class $\mathcal{C}$ given in section \ref{section6}. In the one-out-of-$m$ DQACM protocols of the class $\mathcal{C}$, for $j\in[n]$, the quantum system $A^j=A^j_0A^j_1\cdots A^j_{m-1}$ encodes the $j$th entries $r^j_i\in\Omega$ of the messages $\bold{r}_i$, in a quantum state $\lvert \psi_{r^j}^{s^j}\rangle_{A^j}= \bigotimes_{i\in\text{I}_m}\lvert \alpha^i_{r^j_i}\rangle_{A^j_{s^j_i}}$, where $r^j=(r^j_0,r^j_1,\ldots,r^j_{m-1})$, where the orthogonal bases $\mathcal{D}_i=\{\lvert \alpha^i_{r}\rangle\}_{r\in\Omega}$ are not mutually orthogonal, for $i\in\text{I}_m$, and where $s^j=(s^j_0,s^j_1,\ldots,s^j_{m-1})$ indicates which subsystem of $A^j$ encodes the number $r^j_i$ in the basis $\mathcal{D}_i$, for $i\in\text{I}_m$.

We suggest to extend the class $\mathcal{C}$ to $k$-out-of-$m$ DQACM protocols as follows. For $j\in[n]$, Alice and Bob perform the following actions. In the stage I, Alice randomly chooses $s^j\in\Lambda$ and $r^j_i\in\Omega$, for $i\in\text{I}_m$. Alice prepares $k$ copies of the quantum state $\lvert \psi_{r^j}^{s^j}\rangle_{A^j}$. More precisely, Alice prepares a quantum system $\tilde{A}^j=A^{j,0}A^{j,1}\cdots A^{j,k-1}$, with $A^{j,l}=A^{j,l}_0A^{j,l}_1\cdots A^{j,l}_{m-1}$ for $l\in\text{I}_k$, in the quantum state $\lvert \tilde{\psi}_{r^j}^{s^j}\rangle_{\tilde{A}^j}=\bigotimes_{l\in\text{I}_k}\lvert \psi_{r^j}^{s^j}\rangle_{A^{j,l}}$ that she gives to Bob. Bob measures each subsystem $A^{j,l}_0,A^{j,l}_1,\ldots,A^{j,l}_{m-1}$ of $A^{j,l}$ in the basis $\mathcal{D}_{c_l}$ and obtains classical measurement outcomes $d^{j,l}=(d^{j,l}_0,d^{j,l}_1,\ldots,d^{j,l}_{m-1})$, for $l\in\text{I}_k$, where $c_0,c_1,\ldots,c_{k-1}$ are different numbers input by Bob from the set $\text{I}_m$ indicating that Bob wishes to learn Alice's inputs $\bold{r}_{c_0},\bold{r}_{c_1},\ldots,\bold{r}_{c_{k-1}}$. In the stage II, Alice gives $s^j$ to Bob. Bob then uses $c_0,c_1,\ldots,c_{k-1}$, $s^j$ and his outcome $d^{j,l}$ to obtain $r^j_{c_l}$, for $l\in\text{I}_k$. Thus, we see that Bob obtains $\bold{r}_{c_0},\bold{r}_{c_1},\ldots,\bold{r}_{c_{k-1}}$, as required. We leave as an open problem to investigate whether the $k$-out-of-$m$ DQACM protocols of this class are unconditionally secure against dishonest Bob.


\subsection{Further generalizations of  SCOT}

More generally, we can consider SCOT settings in which Alice inputs some messages in specific regions of spacetime, Bob generates inputs in some regions of spacetime, and Bob obtains some outputs correlated to some of Alice's inputs in some specific regions of spacetime. Alice and Bob previously agree on \emph{spacetime constraints} indicating regions of spacetime where Alice should be unable, or able, to obtain specific information about Bob's inputs; and indicating also regions of spacetime where Bob should be able or unable to obtain specific information about Alice's inputs. The SCOT settings and protocols that we have discussed in this paper are particular examples within this general setting.

One could consider generalizations with more than two parties. Additionally, although we have focused here in output spacetime regions that are pairwise spacelike separated, one could also consider that some output spacetime regions are timelike separated. We expect that in the latter case the security guarantees would be softened.

\section{Discussion}
\label{section9}

In addition to the LODT protocol of Ref. \cite{K11.3} and the one-out-of-two SCOT protocols of Refs. \cite{PG15.1,PGK18.2}, the one-out-of-$m$ SCOT protocols and generalizations presented here are further examples of unconditionally secure spacetime-constrained secure computations. Spacetime-constrained secure computation is a research problem initially outlined by Kent \cite{K11.3}, in which, in addition to the requirements of standard secure computations \cite{Y82}, the inputs and outputs of the computation are restricted to be within constrained regions of spacetime. By definition, in these tasks the inputs and outputs consist in classical information. 

It would be interesting to investigate connections between SCOT and other quantum relativistic cryptographic tasks that have some intrinsically quantum inputs, for example, summoning \cite{K13}, in its various versions \cite{K13,HM16,AK15.3,K18.2,HM18}, where a given quantum state must be returned at specific regions of spacetime. In particular, in the localize-exclude task introduced in Ref. \cite{HM18}, a quantum state must be localized to a collection of authorized spacetime regions while guaranteeing that the state cannot be localized to unauthorized spacetime regions. It would be interesting to investigate connections between SCOT and the localize-exclude task, or other versions of summoning. For example, can SCOT be used as a subroutine to implement a summoning task, or vice versa?

The defined task of one-out-of-$m$ DQACM allowed us to construct an unconditionally secure class $\mathcal{P}_{\text{CC}}$ of one-out-of-$m$ SCOT protocols that do not require to transmit quantum states between distant locations. We provided examples of unconditionally secure one-out-of-$m$ DQACM protocols, hence of unconditionally secure one-out-of-$m$ SCOT protocols. We believe that one-out-of-$m$ DQACM and one-out-of-$m$ SCOT, and generalizations (e.g. in the $k$-out-of-$m$ setting), may be useful primitives to build other cryptographic tasks with no-communication constraints, due to spacelike separation or otherwise. For example, our proposed (or other) unconditionally secure protocols for one-out-of-$m$ DQACM can be used to implement the task of bit string coordination, which is a primitive to perform some supermoney schemes: virtual tokens that are capable to guarantee unconditional security based on the laws of quantum physics and relativity \cite{KSmoney}.

The tasks of one-out-of-$m$ and $k$-out-of-$m$ distributed quantum access with classical memory (DQACM) introduced here seem related to quantum random access codes, with security conditions similar to those of one-out-of-$m$ and $k$-out-of-$m$ oblivious transfer, hence the name we chose to denote these tasks. Broadly speaking, in a quantum random access code (QRAC) \cite{W83,ANTV99}, Alice encodes various classical messages in a quantum state, and Bob decides which message to access. For example, a $(n,m,p)$ quantum random access code is a scheme in which Alice encodes $n$ bits into $m$ qubits in such a way that Bob can recover any bit of his choice with a probability $p>\frac{1}{2}$, where in general one considers $n>m$. The first motivation to study QRACs was given by Wiesner \cite{W83}, who introduced the concept of QRACs with the name of `conjugate coding', in quantum cryptography: quantum money that is impossible to counterfeit. In the literature of QRACs, the questions that are mainly investigated relate to the efficiency of the encodings. For example, one investigates for which values of $m$ and $n$ with $n>m$ there exist $(n,m,p)$ QRACs with $p>\frac{1}{2}$ \cite{ANTV99,ANTV02,HINRY06}, the maximum achievable values of $p$ given $m$ and $n$, how extra resources like randomness \cite{ALMO09} and entanglement \cite{PZ10} improve the efficiency of the encodings, etc. Extensions in which Alice encodes $n$ dits in $m$ qudits, where Bob can retrieve any dit of his choice with probability $p>\frac{1}{d}$, for $d\geq2$, are considered in Ref. \cite{L17}. Extensions of QRACs codes in which Alice encodes, and Bob decodes, intrinsically quantum information were introduced in Ref. \cite{QIC}. In Ref. \cite{SBKTP09}, a variation of QRACs denoted as parity oblivious multiplexing was investigated within a framework of operational theories containing quantum theory as a particular case, and an experimental demonstration of QRACs was performed. It would be interesting to investigate these questions for DQACM, and to investigate further connections between QRACs and DQACM. In particular, can we use results, or intuitions, gained from QRACs to construct unconditionally secure DQACM protocols?

Our proposed protocols for one-out-of-$m$ DQACM and SCOT tolerate small error rates, but they do not consider losses. Although dealing with losses is standard in quantum cryptography, it would be interesting to investigate explicit protocols, as in the lines suggested in Ref. \cite{PGK18.2} for the one-out-of-two case, for instance. Obtaining unconditionally secure protocols with higher allowed error rated would be helpful too. Furthermore, it would be interesting to prove, or disprove, that our proposed $k$-out-of-$m$ DQACM protocols are unconditionally secure. More generally, it would be interesting to find further unconditionally secure SCOT protocols, for example, for the generalized versions of SCOT suggested in section \ref{section8}.

\begin{acknowledgments}
The author acknowledges financial support from the European Research Council project QCC and from the project SPACE17RPSMT-SATT1PITKER during his work at IRIF, Universit\'e Paris Diderot, and from 
the UK Quantum Communications Hub grant no. EP/M013472/1 during his work at the CQIF, DAMTP, University of Cambridge.
\end{acknowledgments}

\appendix

\section{Proof of the bound (\ref{15})}
\label{appendix1}

We show the bound (\ref{15}) for the case $l_0=0$ and $l_1=1$. The proof follows straightforwardly for the general case $l_0,l_1\in\text{I}_m$ with $l_0\neq l_1$.

For fixed $\bold{s}$ and $\bold{v}$, we define the sets $\tau=\bigl\{j\in[n]\big\vert  s^j_{v^j,1}=s^j_0 \bigr\}$ and $\tau_c=\bigl\{j\in[n]\big\vert s^j_{v^j,1}\neq s^j_0\bigr\}$. We define $\omega_{\bold{v}}=\lvert\tau\rvert$, that is, $\omega_{\bold{v}}$ is the number of entries $v^j$ of $\bold{v}$ corresponding to a permutation that takes $(a_0,a_1,\ldots,a_{m-1})\in\Lambda$ to $(?,a_0,?,?,\ldots,?)\in\Lambda$ where `?' denotes any allowed entry after the permutation. As explicitly stated by the notation, we see that $\omega_{\bold{v}}$ only depends on $\bold{v}$, but not on $\bold{s}$. Using the definitions (\ref{x1}) and (\ref{18}), we express $F_{\bold{s}}$ and $G_{\bold{s}_{\bold{v}}}$ by
\begin{eqnarray}
\label{19}
F_{\bold{s}}&=&\sum_{\bold{r} }\Biggl[\bigotimes_{\substack{j\in\tau\\i\in\text{I}_m}}\Bigl( \bigl\lvert \alpha_{r_i^j}^i\bigr\rangle\bigl\langle \alpha_{r_i^j}^i\bigr\rvert\Bigr)_{C_{s_i^j}^j}  \bigotimes_{\substack{j\in\tau_c\\i\in\text{I}_m}}\Bigl( \bigl\lvert \alpha_{r_i^j}^i\bigr\rangle\bigl\langle \alpha_{r_i^j}^i\bigr\rvert\Bigr)_{C_{s_i^j}^j} \times\nonumber\\
&&\qquad\qquad\quad\times\bigotimes\bigl( \Pi^{\bold{r}_0}_{0\bold{s}}\bigr)_{B_0}\bigotimes\mathds{1}_{B_1}\Biggr],\nonumber\\
G_{\bold{s}_{\bold{v}}}&=&\sum_{\bold{r} }\Biggl[\bigotimes_{\substack{j\in\tau\\i\in\text{I}_m}}\Bigl( \bigl\lvert \alpha_{r_i^j}^i\bigr\rangle\bigl\langle \alpha_{r_i^j}^i\bigr\rvert\Bigr)_{C_{s_{v^j,i}^j}^j} \times\nonumber\\
&&\qquad\qquad\quad\times\bigotimes_{\substack{j\in\tau_c\\i\in\text{I}_m}}\Bigl( \bigl\lvert \alpha_{r_i^j}^i\bigr\rangle\bigl\langle \alpha_{r_i^j}^i\bigr\rvert\Bigr)_{C_{s_{v^j,i}^j}^j}  \times\nonumber\\
&&\qquad\qquad\qquad\quad\times\bigotimes \mathds{1}_{B_0}\bigotimes (\Pi^{\bold{r}_{1}}_{1\bold{s}_{\bold{v}}})_{B_1} \biggr].
\end{eqnarray}

Below we compute $F_{\bold{s}}G_{\bold{s}_{\bold{v}}}F_{\bold{s}}$. 
We express the left hand operator $F_{\bold{s}}$ in terms of the dummy variables $\bold{r}=(\bold{r}_0,\bold{r}_1,\ldots,\bold{r}_{m-1})\in\Omega^{nm}$ and the right hand one in terms of $\bold{z}=(\bold{z}_0,\bold{z}_1,\ldots,\bold{z}_{m-1})\in\Omega^{nm}$. The operator $G_{\bold{s}_{\bold{v}}}$ is expressed in terms of $\bold{w}=(\bold{w}_0,\bold{w}_1,\ldots,\bold{w}_{m-1})\in\Omega^{nm}$. For this computation we use the following properties: 1) from the definitions of $\tau$ and $\tau_c$, we have that $s^j_{v^j,1}=s^j_0$ for $j\in\tau$, and $s^j_{v^j,1}\neq s^j_0$ for $j\in\tau_c$; 2) summing over $\bold{z}_0$ we obtain $\bold{z}_0\rightarrow \bold{r}_0$ because $\Pi^{\bold{r}_0}_{0\bold{s}}\Pi^{\bold{z}_0}_{0\bold{s}}=\delta_{\bold{r}_0,\bold{z}_0}\Pi^{\bold{r}_0}_{0\bold{s}}$, since $\bigl\{\Pi^{\bold{r}_0}_{0\bold{s}}\bigr\}_{\bold{r}_0\in\Omega^n}$ is a projective measurement; and 3) $\sum_{a\in\Omega} \lvert \alpha_a^i\rangle\langle \alpha_a^i\rvert=\mathds{1}$ (the identity on a $l$-dimensional Hilbert space) because $\mathcal{D}_i=\{\lvert \alpha_a^i\rangle\}_{a\in\Omega}$ is an orthonormal basis of a $l$-dimensional Hilbert space, for $i\in\text{I}_m$. Thus, after summing over $\bold{r}_i,\bold{w}_{i'},\bold{z}_{i''}$, for $i,i',i''\in\text{I}_{m}$ with $i\neq 0$ and $i'\neq 1$, we obtain
\begin{eqnarray}
\label{20.5}
F_{\bold{s}}G_{\bold{s}_{\bold{v}}}F_{\bold{s}}&=&\sum_{\bold{r}_0,\bold{w}_{1}}
\Biggl[\bigotimes_{j\in\tau}
\biggl(\Bigl\lvert\bigl \langle \alpha_{r_0^j}^0\Big\vert \alpha_{w_{1}^j}^{1}\Bigr\rangle\Bigr\rvert^2 \times\nonumber\\
&&\quad\qquad\times\Bigl( \Bigl\lvert \alpha_{r_0^j}^0\Bigr\rangle\Bigl\langle \alpha_{r_0^j}^0\Bigr\rvert\Bigr)_{C_{s_0^j}^j} \bigotimes_{i\in[m-1]} \mathds{1}_{C_{s_i^j}^j} \biggr)\times\nonumber\\
&&\quad\qquad\times\bigotimes_{j\in\tau_c}\biggl(\Bigl( \Bigl\lvert \alpha_{r_0^j}^0\Bigr\rangle\Bigl\langle \alpha_{r_0^j}^0\Bigr\rvert\Bigr)_{C_{s_0^j}^j} \times\nonumber\\
&&\quad\qquad\times\bigotimes \Bigl(\Bigl\lvert \alpha_{w_{1}^j}^{1}\Bigr\rangle\Bigl\langle \alpha_{w_{1}^j}^{1}\Bigr\rvert\Bigr)_{C_{s_{v^j,1}^j}^j}\times\nonumber\\
&&\qquad\quad\qquad\times\bigotimes_{i\in\text{I}_{v^j}}\mathds{1}_{C_{s^j_i}^j}\biggr)\bigotimes\bigl( \Pi^{\bold{r}_0}_{0\bold{s}}\bigr)_{B_0} \times\nonumber\\
&&\qquad\qquad\quad\qquad\times\bigotimes\bigl( \Pi^{\bold{w}_{1}}_{1\bold{s}_{\bold{v}}}\bigr)_{B_1}\Biggr],
\end{eqnarray}
where $\text{I}_{v^j}=\bigl\{i\in[m-1]\big\vert s_i^j\neq s^j_{v^j,1}\bigr\}$.

Using $\Bigl\lvert\Bigl \langle \alpha_{r_0^j}^0\Big\vert \alpha_{w_{1}^j}^{1}\Bigr\rangle\Bigr\rvert^2\leq \lambda$ from (\ref{x3}) and $\lvert \tau\rvert=\omega_{\bold{v}}$, we obtain
\begin{eqnarray}
\label{21}
F_{\bold{s}}G_{\bold{s}_{\bold{v}}}F_{\bold{s}}&\leq&(\lambda)^{\omega_\bold{v}}\sum_{\bold{r}_0,\bold{w}_{1}}
\Biggl[\bigotimes_{j\in\tau}
\biggl(\Bigl( \Bigl\lvert \alpha_{r_0^j}^0\Bigr\rangle\Bigl\langle \alpha_{r_0^j}^0\Bigr\rvert\Bigr)_{C_{s_0^j}^j} \times\nonumber\\
&&\quad\times\bigotimes_{i\in[m-1]} \mathds{1}_{C_{s_i^j}^j} \biggr)\bigotimes_{j\in\tau_c}\biggl(\Bigl( \Bigl\lvert \alpha_{r_0^j}^0\Bigr\rangle\Bigl\langle \alpha_{r_0^j}^0\Bigr\rvert\Bigr)_{C_{s_0^j}^j} \times\nonumber\\
&&\quad\qquad\times\bigotimes \Bigl(\Bigl\lvert \alpha_{w_{1}^j}^{1}\Bigr\rangle\Bigl\langle \alpha_{w_{1}^j}^{1}\Bigr\rvert\Bigr)_{C_{s^j_{v^j,1}}^j}\times\nonumber\\
&&\qquad\quad\qquad\times\bigotimes_{i\in\text{I}_{v^j}}\mathds{1}_{C_{s_i^j}}\biggr)\bigotimes\bigl( \Pi^{\bold{r}_0}_{0\bold{s}}\bigr)_{B_0} \times\nonumber\\
&&\qquad\qquad\quad\qquad\times\bigotimes\bigl( \Pi^{\bold{w}_{1}}_{1\bold{s}_{\bold{v}}}\bigr)_{B_1}\Biggr].
\end{eqnarray}

Using that $\bigl\{\Pi^{\bold{r}_0}_{0\bold{s}}\bigr\}_{\bold{r}_0\in\Omega^n}$ and $\bigl\{\Pi^{\bold{w}_{1}}_{1\bold{s}_{\bold{v}}}\bigr\}_{\bold{w}_{1}\in\Omega^n}$ are projective measurements, it is straightforward to see that the right-hand term of (\ref{21}) times $(\lambda)^{-\omega_\bold{v}}$ is a projector. Thus, $\lVert (\lambda)^{-\omega_{\bold{v}}}F_{\bold{s}}G_{\bold{s}_{\bold{v}}}F_{\bold{s}}\rVert \leq1$, which implies (\ref{15}).

We have shown (\ref{15}) for the particular case $l_0=0$ and $l_1=1$. But, since from (\ref{x3}) we have $\bigl\lvert\bigl \langle \alpha_{r}^{l_0}\big\vert \alpha_{w}^{l_1}\bigr\rangle\Bigr\rvert^2\leq \lambda$ for any pair of different numbers $l_0,l_1\in\text{I}_m$, and for any $r,\omega\in\Omega$, it is straightforward to see from the derivation above that the bound (\ref{15}) holds for any pair of different numbers $l_0,l_1\in\text{I}_m$.

\section{Details about the quantum measurements of Bob's agents}
\label{appendix2}

Here we show that the following two procedures (1) and (2) described below are mathematically equivalent. More precisely, for any pair of different numbers $i,j\in\text{I}_m$, we show that the joint probability that Bob's agent $\mathcal{B}_i$ obtains a particular outcome $\bold{e}_i$ as his guess of $\bold{r}_{b'-i}$ and Bob's agent $\mathcal{B}_j$ obtains a particular outcome $\bold{e}_j$ as his guess of $\bold{r}_{b'-j}$ in procedure (1) is the same in procedure (2), for any $\bold{e}_i,\bold{e}_j\in\tilde{\Omega}$, where it is assumed that $\Omega_k=\tilde{\Omega}$ is the set of possible values of $\bold{r}_k$, for $k\in I_m$.

In the procedure (1), Bob's agent $\mathcal{B}$ receives the quantum state $\lvert \Psi_{\bold{r}}^{\bold{s}}\rangle_A$ in a quantum system $A$ from Alice's agent $\mathcal{A}$, he introduces an ancillary system $E$ and applies a unitary operation $U$ on $AE$, then he applies a quantum measurement $\tilde{\text{M}}'$ on $AE$ obtaining a classical outcome $(b',i,j)\in\Gamma$, with $(b',i,j)$ being recorded in systems $B_0''$, $B_1''$ and $B''$, where $\Gamma=\{(k,i,j)\in\text{I}_m\times \text{I}_m\times \text{I}_m\vert i\neq j\}$. $\mathcal{B}$ partitions the joint system $AE$ into $B_0$ and $B_1$. $\mathcal{B}$ inspects $(b',i,j)$ from his system $B''$ and then he sends $b'$ to Alice's agent $\mathcal{A}$, and $B_0$ ($B_1$) and $B_0''$ ($B_1''$) to Bob's agent $\mathcal{B}_i$ ($\mathcal{B}_j$). Bob's agent $\mathcal{B}_i$ ($\mathcal{B}_j$) obtains the value $(b',i,j)$ from the system $B_0''$ ($B_1''$) and then applies a projective measurement $\tilde{\text{M}}_{0,b',i,j}^\bold{s}$ ($\tilde{\text{M}}_{1,b',i,j}^\bold{s}$) on $B_0$ ($B_1$) and obtains a classical outcome $\bold{e}_i$ ($\bold{e}_j$) which is his guess of $\bold{r}_{b'-i}$ ($\bold{r}_{b'-j}$).

In the procedure (2), $\mathcal{B}$ applies the following quantum operation $O'$ on $AE'$, where $E'=EB_0''B_1''B''$: $\mathcal{B}$ prepares the quantum system $B_0''B_1''B''$ in a quantum state $\lvert\mu_{0,0,1}\rangle_{B_0''}\otimes\lvert\mu_{0,0,1}\rangle_{B_1''}\otimes\lvert\mu_{0,0,1}\rangle_{B''}$, he applies a unitary operation $U$ on $AE$, and then he applies a unitary operation $U'$ on the total system $B_0B_1B_0''B_1''B''$, where the joint system $AE$ is partitioned into the subsystems $B_0$ and $B_1$ (as in the procedure (1)). The unitary operation $U'$ consists in $\mathcal{B}$ applying the quantum measurement $\tilde{\text{M}}'$ on $B_0B_1$ and preparing each of the quantum systems $B_0''$, $B_1''$ and $B''$ in a quantum state $\lvert \mu_{b',i,j}\rangle$, conditioned on the outcome of $\tilde{\text{M}}'$ being $(b',i,j)$, where $\{\lvert \mu_{b',i,j}\rangle\}_{(b',i,j)\in\Gamma}$ is an orthonormal basis of each of the quantum systems $B_0''$, $B_1''$ and $B''$, for $(b',i,j)\in\Gamma$. Conditioned on the outcome of $\tilde{\text{M}}'$ being $(b',i,j)$, $\mathcal{B}$ sends $b'$ to Alice's agent $\mathcal{A}$ in part of the system $B''$, and  $\mathcal{B}$ sends the joint system $B_0B_0''$ ($B_1B_1''$) to Bob's agent $\mathcal{B}_i$ ($\mathcal{B}_j$). A quantum measurement $\tilde{\text{M}}_{0}^{\bold{s}}$ is applied on the joint system $B_0'=B_0B_0''B''$, with $\mathcal{B}_i$ obtaining the outcome $\bold{e}_i$ from $B_0B_0''$, which is his guess of $\bold{r}_{b'-i}$. Bob's agent $\mathcal{B}_j$ applies a quantum measurement $\tilde{\text{M}}_{1}^{\bold{s}}$ on the joint system $B_1'=B_1B_1''$ and obtains a classical outcome $\bold{e}_j$, which is his guess of $\bold{r}_{b'-j}$.

We give details of the procedures (1) and (2) described above. The quantum state $\lvert\Psi_\bold{r}^\bold{s}\rangle_A$ is transmitted to Bob's agent $\mathcal{B}$. Bob's agent $\mathcal{B}$ introduces an ancillary system $E'$, which includes a system $E$ of arbitrary finite Hilbert space dimension and extra ancillary systems $B_0''$, $B_1''$ and $B''$, each one of Hilbert space dimension $m^2(m-1)$. The system $E$ is set initially to an arbitrary quantum state $\lvert\chi\rangle$, and the systems $B_0''$, $B_1''$ and $B''$ are set initially to the state $\lvert \mu_{0,0,1}\rangle$, where $\{\lvert \mu_{k,i,j}\rangle\vert (k,i,j)\in\Gamma \}$ is an orthonormal basis of $B_0''$, $B_1''$ and $B''$, and where $\Gamma=\{(k,i,j)\in\text{I}_m\times \text{I}_m\times \text{I}_m\vert i\neq j\}$. Bob's agent $\mathcal{B}$ applies an arbitrary unitary operation $U$ on the joint quantum system $AE$. The global state is transformed into the state $\lvert \Phi_\bold{r}^\bold{s}\rangle_{B_0B_1}\lvert \mu_{0,0,1}\rangle_{B_0''}\lvert\mu_{0,0,1}\rangle_{B_1''}\lvert\mu_{0,0,1}\rangle_{B''}$, where
\begin{equation}
\label{app1}
\lvert \Phi_\bold{r}^\bold{s}\rangle_{B_0B_1}=U_{AE}\lvert\Psi_\bold{r}^\bold{s}\rangle_A\lvert\chi\rangle_E,
\end{equation}
and where the joint quantum system $AE$ is partitioned into two subsystems $B_0$ and $B_1$.

Consider the unitary operation $U'$ applied on the whole system $AE'= B_0B_1B_0''B_1''B''$:
\begin{eqnarray}
\label{app2}
U'&=&\sum_{(b',i,j)\in\Gamma}\Bigl[ (R_{b',i,j})_{B_0B_1}\otimes (W_{b',i,j})_{B_0''}\times\nonumber\\
&&\qquad\qquad\times\otimes (W_{b',i,j})_{B_1''}\otimes (W_{b',i,j})_{B''}\Bigr],
\end{eqnarray}
where $W_{b',i,j}$ is a unitary operation  acting on a Hilbert space of dimension $m^2(m-1)$ satisfying $W_{b',i,j}\lvert \mu_{0,0,1}\rangle=\lvert \mu_{b',i,j}\rangle$, for $(b',i,j)\in\Gamma$; and where $\tilde{\text{M}}'=\{R_{b',i,j}\}_{(b',i,j)\in\Gamma}$ is a projective measurement on $B_0B_1$. Consider the projective measurement $\tilde{\text{M}}_{a,b',i,j}^\bold{s}=\{\Pi_{a,\bold{s},b',i,j}^{\bold{e}}\}_{\bold{e}\in\tilde{\Omega}}$ on $B_a$, for $a\in\{0,1\}$, $(b',i,j)\in\Gamma$ and $\bold{s}\in\Lambda_{\text{basis}}$, where $\Lambda_{\text{basis}}$ is the set of possible values of $\bold{s}$.

Consider the projectors
\begin{eqnarray}
\label{app3}
\Pi_{0,\bold{s}}^\bold{e}&=&\sum_{(b',i,j)\in\Gamma}\Bigl[ \bigl(\Pi_{0,\bold{s},b',i,j}^{\bold{e}}\bigr)_{B_0}\otimes\bigl(\lvert \mu_{b',i,j}\rangle\langle \mu_{b',i,j}\rvert \bigr)_{B_0''}\times\nonumber\\
&&\qquad\qquad\times\otimes\mathds{1}_{B''}\Bigr]
\end{eqnarray}
acting on $B_0'=B_0B_0''B''$, and the projectors 
\begin{equation}
\label{app3.1}
\Pi_{1,\bold{s}}^\bold{e}=\sum_{(b',i,j)\in\Gamma} \bigl(\Pi_{1,\bold{s},b',i,j}^{\bold{e}}\bigr)_{B_1}\otimes\bigl(\lvert \mu_{b',i,j}\rangle\langle \mu_{b',i,j}\rvert \bigr)_{B_1''}
\end{equation}
acting on $B_1'=B_1B_1''$, for $\bold{e}\in\tilde{\Omega}$ and $\bold{s}\in\Lambda_{\text{basis}}$. It is straightforward to see that $\tilde{\text{M}}_a^\bold{s}=\{\Pi_{a\bold{s}}^\bold{e}\}_{\bold{e}\in\tilde{\Omega}}$ is a projective measurement acting on $B_a'$, for $a\in\{0,1\}$ and $\bold{s}\in\Lambda_{\text{basis}}$.

It is straightforward to see that, for any pair of different numbers $i,j\in\text{I}_m$, the joint probability that $\mathcal{B}_i$ obtains a particular outcome $\bold{e}_i$ as his guess of $\bold{r}_{b'-i}$ and $\mathcal{B}_j$ obtains a particular outcome $\bold{e}_j$ as his guess of $\bold{r}_{b'-j}$ in procedure (1) is the same in procedure (2), for any $\bold{e}_i,\bold{e}_j\in\tilde{\Omega}$, as claimed.

%

\end{document}